\documentclass[a4paper,11pt]{article}
\usepackage{tensor}
\usepackage{slashed}
\usepackage{jheppub}
\usepackage{lineno}
\usepackage{xcolor}
\usepackage{bm}
\usepackage{mathtools}
\usepackage[compat=1.1.0]{tikz-feynman}
\usepackage{cleveref}
\crefname{equation}{eq.}{eqs.}
\Crefname{equation}{Eq.}{Eqs.}
\crefname{figure}{figure}{figures}
\Crefname{section}{Section}{Sections}
\crefname{appendix}{appendix}{appendices}

\newcommand{\overbar}[1]{\mkern 1.5mu\overline{\mkern-1.5mu#1\mkern-1.5mu}\mkern 1.5mu}
\newcommand{\overlrarrow}[1]{\overset{\text{\tiny$\leftrightarrow$}}{#1}}
\newcommand{\vbar}{\big |}
\newcommand{\biggvbar}{\bigg |}
\DeclareMathOperator{\Tr}{Tr}
\newcommand{\ccomma}{\mathbin{\raisebox{0.5ex}{,}}}

\title{Fermi Geometry of the Higgs Sector}

\author[a,b]{Nathaniel Craig,}
\author[a,c]{I-Kwan Lee,}
\author[a]{and Yu-Tse Lee}

\affiliation[a]{Department of Physics, University of California, Santa Barbara, CA 93106, USA}
\affiliation[b]{Kavli Institute for Theoretical Physics, Santa Barbara, CA 93106, USA}
\affiliation[c]{Department of Physics, National Taiwan University, Taipei 10617, Taiwan}

\emailAdd{ncraig@ucsb.edu}
\emailAdd{b10202045@ntu.edu.tw}
\emailAdd{yutselee@ucsb.edu}

\abstract{
We develop the field space geometry of scalar-fermion effective field theories as a vector bundle supermanifold. We further establish a Fermi normal coordinate system on the bundle that clarifies the geometric content in scattering amplitudes, particularly the imprints of field space non-analyticities. Specializing to the Standard Model Higgs sector, we examine the geometric consequences of custodial symmetry violation, including implications for the physical Higgs field as a distinguished scalar axis and deformations in the fermionic sector. Our results enable a systematic and realistic geometric interpretation of Higgs sector phenomenology.}

\begin{document}
\maketitle
\flushbottom

\section{Introduction}

The discovery of the Higgs boson at the Large Hadron Collider \cite{ATLAS:2012yve, CMS:2012qbp} marks a milestone in the experimental confirmation of the Standard Model (SM), but leaves many pertinent questions like the origin of electroweak symmetry breaking to be answered. With no indication of new massive states on the horizon, the framework of effective field theory (EFT) has emerged as an instrumental tool for interpreting ongoing searches for physics beyond the SM in terms of the accessible degrees of freedom \cite{Brivio:2017vri}. In treating the SM as an EFT, one typically organizes all operators comprising SM fields and consistent with its ultraviolet symmetries by their mass dimensions, with new physics corresponding to Wilson coefficients at higher orders \cite{Weinberg:1979sa, Leung:1984ni, Buchmuller:1985jz}. The resulting Standard Model EFT (SMEFT) is itself a subset of the Higgs EFT (HEFT), organized by derivative power-counting around the infrared symmetries of the Standard Model \cite{Feruglio:1992wf, Bagger:1993zf, Koulovassilopoulos:1993pw}; together they provide a fully generic and model-agnostic low-energy parameterization of the effects of new UV physics on the known SM particles.

However, EFTs generally suffer from an assortment of redundancies that obscure their underlying physical content. Notably, the freedom to perform field redefinitions shuffles contributions between operators with no effect on observables like scattering amplitudes \cite{Kamefuchi:1961sb, Chisholm:1961tha, Coleman:1969sm, Arzt:1993gz}, leaving ambiguity in experimental bounds on the EFT parameter space. A remedy comes from the observation that field redefinitions are akin to coordinate transformations in differential geometry which leave tensors invariant. This analogy has long been appreciated in generic scalar field theories \cite{Meetz:1969as, Honerkamp:1971sh, Honerkamp:1971xtx, Ecker:1971xko, Alvarez-Gaume:1981exa, Alvarez-Gaume:1981exv, Boulware:1981ns, Howe:1986vm, Dixon:1989fj}, and for the problem at hand, reveals that deviations from the SM can be understood as curvature on the space of Higgs field values \cite{Alonso:2015fsp, Alonso:2016btr, Alonso:2016oah, Nagai:2019tgi, Helset:2020yio, Cohen:2020xca, Cohen:2021ucp, Alonso:2021rac, Talbert:2022unj}, and that the suitability of SMEFT or HEFT depends on the symmetries and singularity structure of field space \cite{Alonso:2015fsp, Alonso:2016oah, Cohen:2020xca}. Generalizations of the geometric perspective on EFTs have developed on multiple fronts \cite{Cohen:2022uuw, Cheung:2022vnd, Craig:2023wni, Craig:2023hhp, Alminawi:2023qtf, Cohen:2023ekv, Helset:2024vle, Cohen:2024bml, Lee:2024xqa, Cohen:2024fak}, including particles with spin \cite{Helset:2018fgq, Finn:2019aip, Finn:2020nvn, Helset:2022tlf, Gattus:2023gep, Assi:2023zid, Gattus:2024ird}, and connections have been made with various properties like soft behavior \cite{Cheung:2021yog, Derda:2024jvo, Cohen:2025dex} and renormalization group evolution \cite{Alonso:2022ffe, Helset:2022pde, Assi:2023zid, Jenkins:2023rtg, Jenkins:2023bls, Gattus:2024ird, Li:2024ciy, Aigner:2025xyt, Assi:2025fsm}. Nevertheless, further progress in both the geometric construction and assumed symmetries is required for the framework to be methodically applied to interpreting experimental data. The primary goal of this work is to establish a realistic and systematic geometric treatment of scalar-fermion EFTs to second order in the derivative expansion, with an eye towards phenomenological applications to the Higgs sector. 

In this paper, we re-examine the field space geometry of scalars and fermions, formalizing it as a vector bundle supermanifold and supplying the necessary geometric components in this formalism to ensure field redefinition invariance. We also revisit how field space singularities can be detected at a distance, pointing out the utility of Fermi normal coordinates along geodesics emanating from the perturbative vacuum and constructing them for the geometry at hand. To be clear, the geometry of fermions and Fermi normal coordinates are distinct concepts despite sharing the same eponym, and we develop the latter for scalar as well as fermion fields. Thus equipped, we specialize to the SM Higgs sector and study the geometric information that can be recovered from scattering amplitudes. We extend previous work \cite{Alonso:2016oah, Cohen:2021ucp} via a more complete treatment of custodial symmetry violation, fully generalizing the base space which all other geometric constructions depend upon, and generating noteworthy features in new directions. All in all, this work strives to offer a more comprehensive account of fermionic field space geometry and its application to various aspects of electroweak physics. Nevertheless, we are limited in scope to a field content of scalars and fermions, with longitudinal modes of gauge bosons accommodated by means of the Goldstone boson equivalence theorem. We primarily focus on one generation of quarks, although there is no conceptual challenge in extending to leptons or multiple flavors. We work at energies no more than the EFT cutoff, and for the most part ignore the effects of color or loops.

The rest of the paper is organized as follows. \Cref{sec:geometry} lays out the geometric formalism for generic fermionic EFTs and the construction of Fermi normal coordinates in the scalar and fermion sectors. \Cref{sec:application} details the geometry of the SM Higgs sector and significant features at or away from the vacuum that register in scattering amplitudes. \Cref{sec:conclusion} concludes the paper and discusses avenues for future work. Expressions for phase space integrals and various geometric quantities are reserved for a pair of appendices.

\section{Fermions and Fermi normal coordinates}
\label{sec:geometry}

\begin{figure}[tbp]
    \centering
    \includegraphics[width=0.9\columnwidth]{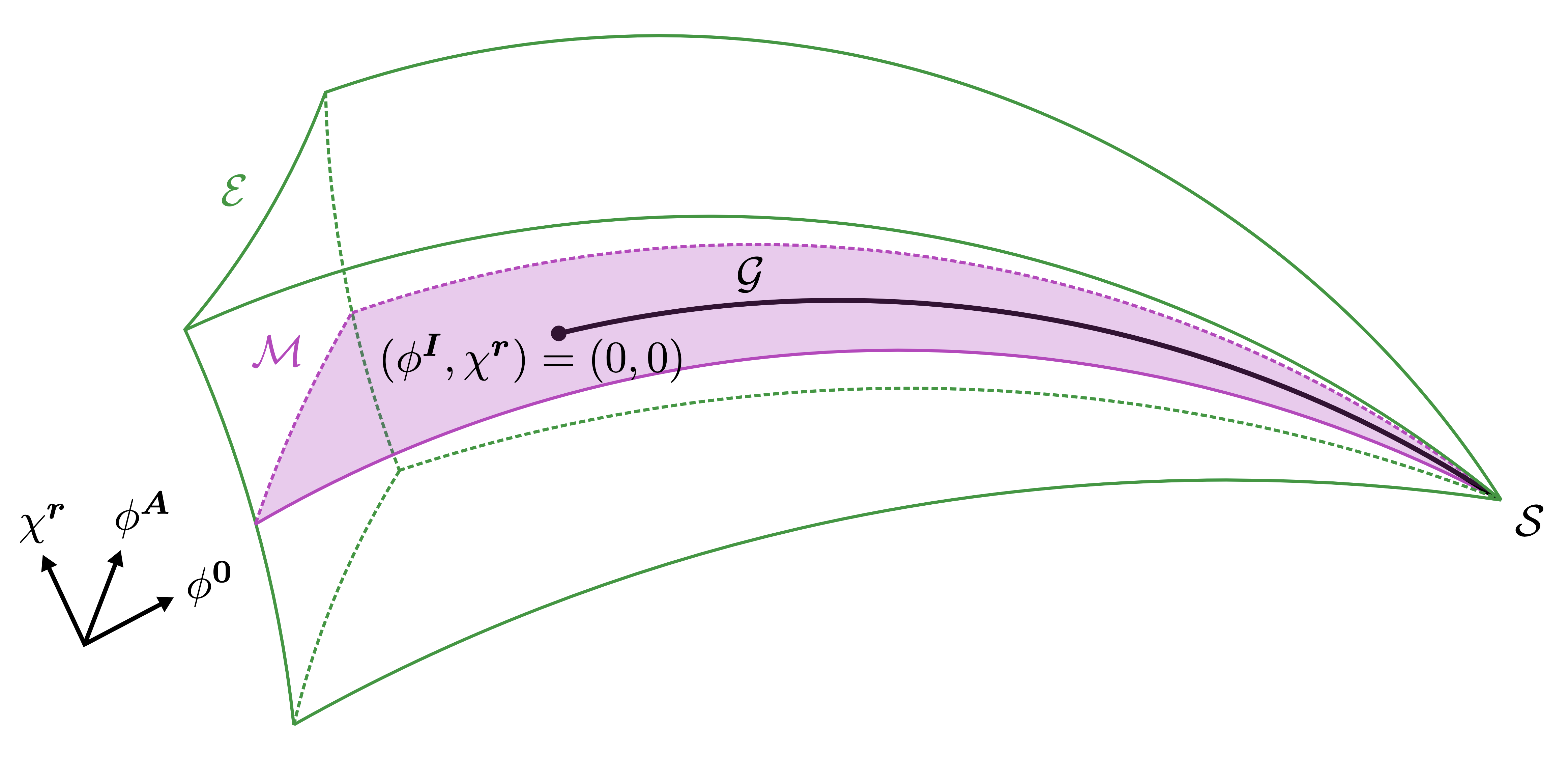}
    \caption{\label{fig:fermi_geometry}An illustration of the vector bundle field space $\mathcal{E}$ of scalars and fermions in Fermi normal coordinates $(\phi^{\bm{I}}, \chi^{\bm{r}})$. The scalar field space constitutes the base space $\mathcal{M}$ over which a complex vector space $\mathcal{V}$ for fermion flavor is bundled. A connection $\Gamma$ specifies how parallel transport is performed between fibers over nearby points and makes $\mathcal{E}$ curved over $\mathcal{M}$. The Fermi normal coordinate system is determined by connecting the origin to another point of interest, such as a singularity (denoted here by $\mathcal{S}$), using a geodesic $\mathcal{G}$ which becomes the axis for the primary coordinate $\phi^{\bm{0}}$. In these coordinates, the geodesic is locally flat and an ending sequence of covariant derivatives in $\phi^{\bm{0}}$ is equivalent to partial ones.}
\end{figure}

Before diving into any specific theory, we reconsider the geometric formulation of generic fermionic EFTs and develop new tools that more closely relate the geometry to observables like scattering amplitudes. We make a principled choice to combine the scalars and fermions into a vector bundle and examine the underlying geometry in detail. When zoomed out to its total space, the construction intersects with the results of previous works to varying degrees \cite{Finn:2020nvn, Assi:2023zid, Derda:2024jvo}, but differs in some structures used for field redefinition invariance. Among the various possible field parameterizations, Riemann normal coordinates have been the system of choice in the literature, \emph{e.g.} \cite{Cohen:2021ucp, Alonso:2022ffe, Helset:2022pde, Assi:2023zid}, to compute amplitudes and loop corrections using the geometric framework. We identify and construct an alternative system in both the scalar and fermion sectors, namely Fermi normal coordinates \cite{Fermi:1922sopra, Fermi:1922sopra2}, that as the subsequent section will demonstrate are more suitable for the purpose of probing geometric features beyond the perturbative vacuum. The geometric constructions are schematically depicted in \cref{fig:fermi_geometry}.

\subsection{Formulating field space as a vector bundle}

The field space geometry of an EFT is defined by the field redefinitions it permits. For scalar theories, any invertible analytic redefinition
\begin{equation}
    \phi^I = \phi^I(\phi^{J'}) \, ,
\end{equation}
is allowed. These are the transition maps of a real manifold $\mathcal{M}$, which we call the scalar field space. Meanwhile, fermions $\psi^r$ have more structure that redefinitions should respect --- they constitute complex vector spaces of Grassmann spinors.\footnote{So long as a consistent choice is made, the type of spinors does not generally matter as our geometry extracts only the flavor degrees of freedom. We choose to work with Dirac spinors exclusively, projecting onto left- and right-handed spinors when necessary.} In particular, a complex linear combination of differently flavored spinors remains a spinor, so multiplication by an invertible $\phi$-dependent complex matrix $\tau$ is a valid fermion redefinition. We incorporate this redefinition geometrically using a complex vector bundle $\mathcal{E}$ over $\mathcal{M}$, with a fiber $\mathcal{V}$ coordinated by anti-commuting complex numbers $\chi^r$. The transition maps
\begin{equation}
    \chi^r = \tensor{\tau}{^r_{t'}}(\phi) \, \chi^{t'} \, , \label{eq:fermion_redef}
\end{equation}
of $\mathcal{E}$ implement the fermion redefinition, and $\mathcal{V}$ constitutes a fermion flavor space. The spinors $\psi^r$ are then valued in $S \otimes \mathcal{V}$ where $S$ is the appropriate spin representation. \Cref{eq:fermion_redef} spells out what underlies the common notation $\psi^r = \tensor{\tau}{^r_{t'}}(\phi) \, \psi^{t'}$, which we do use in later sections.

Geometric quantities on $\mathcal{E}$ are covariant under the scalar and fermion field redefinitions above, and can be extracted from the Lagrangian upon organizing the operators according to their Lorentz index structures. Consider a generic two-derivative theory\footnote{We count each fermion bilinear as one derivative since it contributes one power of momentum to the scattering amplitude.}
\begin{subequations} \label{eq:Lagrangian}
\begin{align}
    \mathcal{L} &\supset \frac{1}{2} \, g_{IJ}(\phi) \, \partial_\mu \phi^I \partial^\mu \phi^J + \frac{i}{2} \, k_{\bar{p}r}(\phi) \, \overbar{\psi}^{\bar{p}} \overlrarrow{\slashed{\partial}} \psi^r + i \, \omega_{\bar{p}rI}(\phi) \, \overbar{\psi}^{\bar{p}} \gamma^\mu \psi^r \, \partial_\mu \phi^I \label{eq:kinetic} \\
    &\quad - V(\phi) - \overbar{\psi}^{\bar{p}} \, M_{\bar{p}r}(\phi) \, \psi^r + c_{\bar{p}r\bar{s}t}(\phi) \, \overbar{\psi}^{\bar{p}} \gamma^\mu \psi^r \, \overbar{\psi}^{\bar{s}} \gamma_\mu \psi^t + \ldots \, . \label{eq:potential}
\end{align}
\end{subequations}
where $\overbar{\psi}^{\bar{p}} \overlrarrow{\slashed{\partial}} \psi^r = \overbar{\psi}^{\bar{p}} \gamma^\mu \partial_\mu \psi^r - (\partial_\mu \overbar{\psi}^{\bar{p}}) \gamma^\mu \psi^r$ and the ellipsis represents other two- and four-fermion terms. Among the various operators, the kinetic terms in \cref{eq:kinetic} enjoy a special status:
\begin{itemize}
    \item The scalar kinetic term $g_{IJ}$ defines a positive-definite symmetric bilinear form in the scalar directions and yields a Riemannian metric on $\mathcal{M}$. Associated quantities include the Levi-Civita connection $\tensor{\Gamma}{^K_{IJ}}$ on $T\mathcal{M}$ and the Riemann curvature tensor $\tensor{R}{^I_{JKL}}$.
    \item The fermion kinetic term $k_{\bar{p}r}$ defines a positive-definite Hermitian form on the fermion flavor space and yields a Hermitian metric on $\mathcal{E}$.
    \item The fermion kinetic term $\omega_{\bar{p}rI}$ is anti-Hermitian and the combinations
    \begin{equation}
        \tensor{\Gamma}{^p_{Ir}} = k^{p\bar{s}} \left ( \frac{1}{2} \, k_{\bar{s}r, \, I} + \omega_{\bar{s}rI} \right ) \enspace \text{and} \enspace \tensor{\Gamma}{^{\bar{p}}_{I\bar{r}}} = k^{s\bar{p}} \left ( \frac{1}{2} \, k_{\bar{r}s, \, I} - \omega_{\bar{r}sI} \right ) = (\tensor{\Gamma}{^p_{Ir}})^* \, , \label{eq:vec_bundle_conn}
    \end{equation}
    yield a vector bundle connection on $\mathcal{E}$. Here, the comma denotes a partial derivative and $k^{p\bar{s}}$ is the matrix inverse of $k_{\bar{s}p}$, both of which will subsequently be used to raise and lower fermion indices.
\end{itemize}
Equivalently, the tangent vectors
\begin{equation}
    \frac{\delta}{\delta \phi^I} = \frac{\partial}{\partial \phi^I} - \tensor{\Gamma}{^p_{Ir}} \, \chi^r \frac{\partial}{\partial \chi^p} - \tensor{\Gamma}{^{\bar{p}}_{I\bar{r}}} \, \overbar{\chi}^{\bar{r}} \frac{\partial}{\partial \overbar{\chi}^{\bar{p}}} \, ,
\end{equation}
span a horizontal subbundle of $T\mathcal{E}$, \emph{i.e.} they transform among themselves under \cref{eq:fermion_redef}. Without the connection, the partial derivatives $\partial / \partial \phi^I$ are not tensorial on their own. Meanwhile, $\partial / \partial \chi^r$ and $\partial / \partial \overbar{\chi}^{\bar{p}}$ are tensorial and each spans a vertical subbundle of $T\mathcal{E}$. The dual subbundles in $T^*\mathcal{E}$ are spanned by
\begin{equation}
    d\phi^I \, , \quad \delta \chi^r = d\chi^r + \tensor{\Gamma}{^r_{Ip}} \, \chi^p \,  d\phi^I \quad \text{and} \quad \delta \overbar{\chi}^{\bar{p}} = d\overbar{\chi}^{\bar{p}} + \tensor{\Gamma}{^{\bar{p}}_{I\bar{r}}} \, \overbar{\chi}^{\bar{r}} \, d\phi^I \, .
\end{equation}
Like the Levi-Civita connection on $T\mathcal{M}$, the connection on $\mathcal{E}$ has curvature. The components are given by\footnote{Round (square) brackets denote index (anti-)symmetrization, and vertical bars demarcate segments excluded from (anti-)symmetrization. The (anti-)symmetrization of $n$ indices is normalized by $1/n!$.}
\begin{align}
    \tensor{R}{^p_{rIJ}} &= - \, 2 \, \tensor{\Gamma}{^p_{[I|r,|J]}} + 2 \, \tensor{\Gamma}{^p_{[I|s}} \, \tensor{\Gamma}{^{\vphantom{p}s}_{|J]r}} \, , \\
    \tensor{R}{^{\bar{p}}_{\bar{r}IJ}} &= - \, 2 \, \tensor{\Gamma}{^{\bar{p}}_{[I|\bar{r},|J]}} + 2 \, \tensor{\Gamma}{^{\bar{p}}_{[I|\bar{s}}} \, \tensor{\Gamma}{^{\vphantom{\bar{p}}\bar{s}}_{|J]\bar{r}}} = (\tensor{R}{^p_{rIJ}})^* \, .
\end{align}

Most existing works operate not within the vector bundle $\mathcal{E}$ but rather on its total space. While redundant, we can for the sake of comparison opt to merge all above structures into one by defining a graded Riemannian metric
\begin{equation}
    G = d\phi^I \, g_{IJ} \, d\phi^J + \delta \chi^r \, k_{\bar{p}r} \, \delta \overbar{\chi}^{\bar{p}} - \delta \overbar{\chi}^{\bar{p}} \, k_{\bar{p}r} \, \delta \chi^r \, , \label{eq:graded_metric}
\end{equation}
on the total space. Unlike the three individual kinetic terms, this metric runs over all scalar and fermionic coordinates, and in matrix form reads\footnote{For ease of comparison with literature, we have chosen to express the matrix in the non-tensorial basis $\{ d\phi^I, d\chi^r, d\overbar{\chi}^{\bar{p}} \}$ of $T^*\mathcal{E}$. We follow the Grassmann index conventions of \cite{DeWitt:2012mdz}. The index $c$ runs over scalar indices $I$ and fermion indices $p$ and $\bar{p}$. Likewise, $d$ runs over $J$, $r$ and $\bar{r}$. For later use, we reserve $t$ and $b$ for the fermion indices representing the top and bottom quarks.}
\begin{equation}
    \tensor[_{c\,}]{G}{_d} = 
    \begingroup
    \setlength\arraycolsep{6pt}
    \renewcommand\arraystretch{1.4}
    \begin{pmatrix}
        g_{IJ} + 2 \, \tensor{\Gamma}{^q_{(I|s}} \, \tensor{\Gamma}{_{q|J)\bar{s}}} \, \chi^s \overbar{\chi}^{\bar{s}} & - \, \tensor{\Gamma}{_{rI\bar{s}}} \, \overbar{\chi}^{\bar{s}} & \tensor{\Gamma}{_{\bar{r}Is}} \, \chi^s \\
        \tensor{\Gamma}{_{pJ\bar{s}}} \, \overbar{\chi}^{\bar{s}} & 0 & k_{\bar{r}p} \\
        - \, \tensor{\Gamma}{_{\bar{p}Js}} \, \chi^s & - \, k_{\bar{p}r} & 0 \\
    \end{pmatrix}
    \endgroup
    \, .
\end{equation}
The scalar components on the upper left receive Grassmann corrections from $\delta \chi^r$ and $\delta \overbar{\chi}^{\bar{p}}$, which are needed to ensure that $G$ is tensorial. The presence of an $\mathcal{O}(\chi^2)$ term in $\tensor[_{I\,}]{G}{_J}$ is similar to \emph{e.g.} \cite{Finn:2020nvn, Gattus:2023gep} but differs from \emph{e.g.} \cite{Assi:2023zid, Derda:2024jvo, Assi:2025fsm}.\footnote{When counting Grassmann order like in $\mathcal{O}(\chi^2)$, we include powers of both $\chi^r$ and $\overbar{\chi}^{\bar{p}}$.} On the largest bundle $T\mathcal{E}$, there is then a symmetric $G$-compatible connection with the Christoffel symbols\footnote{The factor $(-1)^c$ represents the Grassmann parity of $c$. Indices are shifted according to $G_{cd} = (-1)^c \tensor[_c]{G}{_d}$. Index symmetry on a supermanifold amounts to $\tensor{\Gamma}{^c_{de}} = (-1)^{de} \, \tensor{\Gamma}{^c_{ed}}$, where the minus sign turns on when both $d$ and $e$ are Grassmann-odd.}
\begin{equation}
    \tensor{\Gamma}{^c_{de}} = \frac{(-1)^{f}}{2} \, G^{cf} \left ( G_{fd,e} + (-1)^{de} \, G_{fe,d} - (-1)^{f(d+e)} \, G_{de,f} \right ) \, .
\end{equation}
If we restrict to the base space $\chi^p = 0$, the non-vanishing symbols are $\tensor{\Gamma}{^K_{IJ}}$, $\tensor{\Gamma}{^p_{Ir}}$ and $\tensor{\Gamma}{^{\bar{p}}_{I\bar{r}}}$. They reproduce the connection coefficients on the smaller bundles $T\mathcal{M}$ and $\mathcal{E}$. Similarly, the curvature tensor
\begin{equation}
    \tensor{R}{^c_{def}} = - \tensor{\Gamma}{^c_{de,f}} + (-1)^{ef} \, \tensor{\Gamma}{^c_{df,e}} + (-1)^{e(a+d)} \, \tensor{\Gamma}{^c_{ae}} \, \tensor{\Gamma}{^a_{df}} - (-1)^{f(a+d+e)} \, \tensor{\Gamma}{^c_{af}} \, \tensor{\Gamma}{^a_{de}} \, ,
\end{equation}
reproduces $\tensor{R}{^I_{JKL}}$, $\tensor{R}{^p_{rIJ}}$ and $\tensor{R}{^{\bar{p}}_{\bar{r}IJ}}$ up to $\mathcal{O}(\chi^2)$. There is no new information in $G$, but we will occasionally use it as a bookkeeping device that subsumes the geometry in the scalar and fermionic directions.

There are other operators of interest that are embedded in separate tensors on field space. In particular, the scalar potential and fermion mass matrix in \cref{eq:potential} are needed to account for the masses of the fields. The potential $V$ is a scalar on $\mathcal{M}$ and the mass matrix $M$ is a tensor on $\mathcal{E}$. Taking $\mathrm{argmin} \, V(\phi) = 0$, we call the point $\phi^I = 0$ the vacuum on $\mathcal{M}$ and $(\phi^I, \chi^r) = (0, 0)$ the vacuum on $\mathcal{E}$. Four-fermion operators such as $c$ are each also a tensor on $\mathcal{E}$. They play a limited role in the vector bundle geometry of $\mathcal{E}$, which is our main interest, but can be incorporated in $G$ to generate new four-fermion curvature components like $\tensor{R}{^p_{r\bar{s}t}}$ on the total space, by \emph{e.g.} modifying $k_{\bar{p}r}$ with $(c_{\bar{p}r\bar{s}t} + c_{\bar{p}t\bar{s}r}) \, \chi^t \overbar{\chi}^{\bar{s}}$ in \cref{eq:graded_metric}.

\subsection{Charting field space with geodesics}

In practice, the field space geometry of an EFT can be reconstructed via the measurement of physical observables --- field redefinition invariance entails that they can be expressed perturbatively in terms of tensors evaluated at the vacuum. While this gives us immediate access to a single point on the manifold, there also exist important geometric features that reside elsewhere \cite{Alonso:2015fsp, Cohen:2020xca}. A good choice of coordinate system goes a long way in mapping out these features. At a technical level, distant geometric information is encoded in an all-order Taylor expansion at the vacuum, but can be obscured by the curvature in between. Fermi normal coordinates provide a computational basis that is locally flat along geodesics from the vacuum to other points of interest \cite{Manasse:1963zz, Li:1979-1, Li:1979-2}, thereby rendering the local decoding of distant features maximally transparent.\footnote{Fermi normal coordinates arise frequently in the context of general relativity; see \emph{e.g.} \cite{Cooperstock:1998ny, Baldauf:2011bh} for an example in a closely related spacetime geometry.} As a first step, we proceed in a general setting to apply this coordinate system to the scalar fields on the base space $\mathcal{M}$, and provide an original construction for the fermionic fields on the vector bundle $\mathcal{E}$.

We start on $\mathcal{M}$ and take a point, \emph{e.g.} the vacuum, to be the origin. Suppose another point of interest sits on a geodesic $\mathcal{G}$ emanating from the origin. We can arrange $\mathcal{G}$ to lie along a primary coordinate $\phi^{\bm{0}}$ that traces the geodesic at unit speed. The tangent vector $\partial_{\bm{0}}$ is covariantly constant along the geodesic, and can be completed into an orthonormal frame $\{ \partial_{\bm{0}}, \partial_{\bm{1}}, \partial_{\bm{2}}, \, \ldots \}$ by choosing a set of complementary vectors at the origin and performing parallel transport along $\mathcal{G}$. Now at each point on the primary geodesic $\mathcal{G}$, we send out an orthogonal secondary geodesic specified by the initial velocity $\phi^{\bm{1}} \, \partial_{\bm{1}} + \phi^{\bm{2}} \, \partial_{\bm{2}} + \ldots$ and label its location after unit time as $\phi^{\bm{I}} = (\phi^{\bm{0}}, \phi^{\bm{A}}) = (\phi^{\bm{0}}, \phi^{\bm{1}}, \phi^{\bm{2}}, \ldots)$. These numbers constitute a Fermi normal coordinate system in a neighborhood of $\mathcal{G}$ and will be distinguished by bolded indices, with uppercase Latin indices near the start of the alphabet representing secondary coordinates.

With the semi-colon indicating a covariant derivative, Fermi normal coordinates allow us to convert between non-covariant and covariant quantities associated with the Levi-Civita connection as follows \cite{Ishii:2005xq}:
\begin{alignat}{3}
    &&\partial^0: \quad &g_{\bm{I} \bm{J}} \vbar_{\mathcal{G}} &&= \, \delta_{\bm{I} \bm{J}} \, , \label{eq:Fermi_identities_0} \\
    &&\partial^1: \quad &\tensor{\Gamma}{^{\bm{I}}_{\bm{J} \bm{K}}} \vbar_{\mathcal{G}} &&= \, 0 \, , \label{eq:Fermi_identities_1} \\
    &&\partial^2: \quad &\tensor{\Gamma}{^{\bm{I}}_{\bm{J} \bm{0}, \, \bm{A}}} \vbar_{\mathcal{G}} &&= \, - \, \tensor{R}{^{\bm{I}}_{\bm{J} \bm{0} \bm{A}}} \vbar_{\mathcal{G}} \, , \label{eq:Fermi_identities_2} \\
    & &&\tensor{\Gamma}{^{\bm{I}}_{\bm{A} \bm{B}, \, \bm{C}}} \vbar_{\mathcal{G}} &&= \, - \, \frac{2}{3} \, \tensor{R}{^{\bm{I}}_{(\bm{A} \bm{B}) \bm{C}}} \vbar_{\mathcal{G}} \, , \\
    &&\partial^3: \quad &\tensor{\Gamma}{^{\bm{I}}_{\bm{0} \bm{0}, \, \bm{A} \bm{B}}} \vbar_{\mathcal{G}} &&= \, - \, \tensor{R}{^{\bm{I}}_{\bm{0} \bm{0} (\bm{A}; \, \bm{B})}} \vbar_{\mathcal{G}} + \tensor{R}{^{\bm{I}}_{(\bm{A} \bm{B}) \bm{0}; \, \bm{0}}} \vbar_{\mathcal{G}} \, , \\
    & &&\tensor{\Gamma}{^{\bm{I}}_{\bm{A} \bm{0}, \, \bm{B} \bm{C}}} \vbar_{\mathcal{G}} &&= \, - \, \tensor{R}{^{\bm{I}}_{\bm{A} \bm{0} (\bm{B}; \, \bm{C})}} \vbar_{\mathcal{G}} + \frac{1}{3} \, \tensor{R}{^{\bm{I}}_{(\bm{B} \bm{C}) \bm{A}; \, \bm{0} }} \vbar_{\mathcal{G}} \, , \\
    & &&\tensor{\Gamma}{^{\bm{I}}_{\bm{A} \bm{B}, \, \bm{C} \bm{D}}} \vbar_{\mathcal{G}} &&= \, - \frac{5}{6} \, \tensor{R}{^{\bm{I}}_{(\bm{A} \bm{B}) (\bm{C}; \, \bm{D})}} \vbar_{\mathcal{G}} + \frac{1}{6} \, \tensor{R}{^{\bm{I}}_{(\bm{C} \bm{D}) (\bm{A}; \, \bm{B})}} \vbar_{\mathcal{G}} \, , \label{eq:Fermi_identities_3} \\
    & && &&\hspace{0.52em} \vdots \notag
\end{alignat}
The formulas have been listed in increasing derivative order in $\phi^{\bm{A}}$. Meanwhile, each of them is valid along the entire line $\mathcal{G}$. In particular, \cref{eq:Fermi_identities_1}, \emph{i.e.} local flatness along $\mathcal{G}$, implies that an ending sequence of covariant derivatives in $\phi^{\bm{0}}$ on any tensor $T$ is equivalent to partial ones when restricted to $\mathcal{G}$. This is evident from the general formula \cite{Hatzinikitas:2000xe}
\begin{align}
    &T_{\bm{I_1} \, \cdots \, \bm{I_M} ; \, \bm{J_1} \, \cdots \, \bm{J_N}} = T_{\bm{I_1} \, \cdots \, \bm{I_M} , \, \bm{J_1} \, \cdots \, \bm{J_N}} \label{eq:cov_deriv} \\
    &\hspace{5em} - \left [ T_{\bm{K} \bm{I_2} \, \cdots \, \bm{I_M}} \, \tensor{\Gamma}{^{\bm{K}}_{\bm{J_1} \bm{I_1}}} + \ldots + T_{\bm{I_1} \, \cdots \, \bm{I_{M-1}} \bm{K}} \, \tensor{\Gamma}{^{\bm{K}}_{\bm{J_1} \bm{I_M}}} \right ]_{, \, \bm{J_2} \, \cdots \, \bm{J_N}} \notag \\
    &\hspace{5em} - \left [ T_{\bm{K} \bm{I_2} \, \cdots \, \bm{I_M} ; \, \bm{J_1}} \, \tensor{\Gamma}{^{\bm{K}}_{\bm{J_2} \bm{I_1}}} + \ldots + T_{\bm{I_1} \, \cdots \, \bm{I_M} ; \, \bm{K}} \, \tensor{\Gamma}{^{\bm{K}}_{\bm{J_2} \bm{J_1}}} \right ]_{, \, \bm{J_3} \, \cdots \, \bm{J_N}} \notag \\
    &\hspace{5.47em} \vdots \notag \\
    &\hspace{5em} - \left [ T_{\bm{K} \bm{I_2} \, \cdots \, \bm{I_M} ; \, \bm{J_1} \, \cdots \, \bm{J_{N-1}}} \, \tensor{\Gamma}{^{\bm{K}}_{\bm{J_N} \bm{I_1}}} + \ldots + T_{\bm{I_1} \, \cdots \, \bm{I_M} ; \, \bm{J_1} \, \cdots \, \bm{J_{N-2}} \bm{K}} \, \tensor{\Gamma}{^{\bm{K}}_{\bm{J_N} \bm{J_{N-1}}}} \right ] \, , \notag
\end{align}
upon taking $\bm{J_1} = \ldots = \bm{J_N} = \bm{0}$. Hence, the conversions \crefrange{eq:Fermi_identities_0}{eq:Fermi_identities_3} continue to apply when differentiated to arbitrary order in $\phi^{\bm{0}}$, whether partially or covariantly on the right hand sides. Crucially, the equivalence of covariant and partial derivatives in $\phi^{\bm{0}}$ relates physical observables directly to Taylor expansions at the vacuum, whose properties can then be deciphered using analytic techniques on Euclidean space.

Note that this coordinate system is different from the commonly used Riemann normal coordinates in \textit{e.g.} \cite{Honerkamp:1971sh, Alonso:2015fsp, Jenkins:2023bls, Li:2024ciy, Aigner:2025xyt}. Riemann normal coordinates would instead be determined solely by geodesics from a single point, \emph{i.e.} the origin, and treat each direction on an equal footing. The analogs of \crefrange{eq:Fermi_identities_0}{eq:Fermi_identities_3} would not discriminate between $\phi^{\bm{0}}$ and $\phi^{\bm{A}}$, but would only hold at the origin. Importantly, the conversion from covariant to partial derivatives produces additional terms of local curvature between the origin and the point of interest, obscuring the interpretation of distant geometry. To be clear, physical observables do not depend on the choice of field space coordinates; what Fermi normal coordinates buys us is a re-organization of contributions within the same observable, such that the geometric significance of each piece becomes most transparent.

We have now established Fermi normal coordinates in the scalar sector and seen how they simplify the functional form of the Levi-Civita connection on the tangent bundle $T\mathcal{M}$. Next in line is the fermion sector, in which analogous simplifications can also be achieved for the connection on the vector bundle $\mathcal{E}$. We proceed to construct a set of fermion coordinates $\psi^{\bm{r}}$ that satisfy the identities
\begin{equation}
    k_{\bm{\bar{p}} \bm{r}} = \delta_{\bm{p} \bm{r}} \, , \quad \omega_{\bm{\bar{p}} \bm{r} \bm{0}} \vbar_{\mathcal{G}} = 0 \, \quad \text{and} \quad \omega_{\bm{\bar{p}} \bm{r} (\bm{A_1}, \, \cdots \, \bm{A_N})} \vbar_{\mathcal{G}} = 0 \quad \text{for all } N \geq 1 \, . \label{eq:k_omega_Fermi_identities}
\end{equation}
In particular, the fermions are canonically normalized and the connection coefficients
\begin{equation}
    \tensor{\Gamma}{^{\bm{p}}_{\bm{I} \bm{r}}} \vbar_{\mathcal{G}} = 0 \quad \text{and} \quad \tensor{\Gamma}{^{\bm{\bar{p}}}_{\bm{I} \bm{\bar{r}}}} \vbar_{\mathcal{G}} = 0 \, ,
\end{equation}
vanish on the line $\mathcal{G}$, which are the analogs of \crefrange{eq:Fermi_identities_0}{eq:Fermi_identities_1}. Since the construction is new, we provide a nuts-and-bolts prescription.

Start in scalar Fermi normal coordinates but arbitrary fermion coordinates. The strategy is to simplify the fermion kinetic terms using their transformation laws
\begin{align}
    k_{\bar{s}' t'} &= \tensor{\overbar{\tau}}{^{\bar{p}}_{\bar{s}'}} \, k_{\bar{p}r} \, \tensor{\tau}{^{\vphantom{\bar{p}} r}_{t' \vphantom{\bar{s}'}}} \, , \\
    \omega_{\bar{s}' t' \bm{I}} &= \tensor{\overbar{\tau}}{^{\bar{p}}_{\bar{s}'}} \, \omega_{\bar{p} r \bm{I}} \, \tensor{\tau}{^{\vphantom{\bar{p}} r}_{t' \vphantom{\bar{s}'}}} - \frac{1}{2} \, \tensor{\overbar{\tau}}{^{\bar{p}}_{\bar{s}', \, \bm{I}}} \, k_{\bar{p}r} \, \tensor{\tau}{^{\vphantom{\bar{p}} r}_{t' \vphantom{\bm{I}}}} + \frac{1}{2} \, \tensor{\overbar{\tau}}{^{\bar{p}}_{\bar{s}' \vphantom{\bm{I}}}} \, k_{\bar{p}r} \, \tensor{\tau}{^{\vphantom{\bar{p}} r}_{t', \, \bm{I}}} \, .
\end{align}
under fermion redefinitions of the form \cref{eq:fermion_redef}. Two successive redefinitions are needed. The Hermitian matrix $k_{\bar{p}r}^{\vphantom{*}} = k_{\bar{r}p}^*$ in fermion indices satisfies
\begin{equation}
    k_{\bar{p}r} = \tensor{ [ \overbar{(\tau^{(k)})^{-1}} ] }{^{\bar{\mathtt{s}}}_{\bar{p}}} \; \delta_{\bar{\mathtt{s}} \mathtt{t}} \; \tensor{ [ (\tau^{(k)})^{-1} ] }{^{\vphantom{\bar{\mathtt{s}}} \mathtt{t}}_{\vphantom{\bar{p}} r}} \, ,
\end{equation}
for some invertible $\tau^{(k)}$ and can be trivialized by first redefining $\psi^r = \tensor{(\tau^{(k)})}{^r_{\mathtt{t}}} \, \psi^{\mathtt{t}}$, yielding the foremost identity in \cref{eq:k_omega_Fermi_identities}. Preserving canonical normalization, there is residual freedom for a second redefinition $\psi^{\mathtt{r}} = \tensor{(\tau^{(\omega)})}{^{\mathtt{r}}_{\bm{t}}} \, \psi^{\bm{t}}$ by a unitary $\tau^{(\omega)} = \exp \Omega$, which we now use to simplify the anti-Hermitian $\omega_{\bar{\mathtt{p}} \mathtt{r} \bm{I}}^{\vphantom{*}} = - \, \omega_{\bar{\mathtt{r}} \mathtt{p} \bm{I}}^*$ along $\mathcal{G}$ so that it yields the remaining identities.

According to the matrix formulas
\begin{align}
    e^{-\Omega} \, \frac{\partial}{\partial \phi^{\bm{I}}} \, e^\Omega &= \sum_{N \geq 0} \frac{(-1)^N}{(N+1)!} \bigg [ \underbrace{\Omega \, \ccomma \, \cdots \, \bigg [ \Omega}_{N \text{ copies}} \, \ccomma \, \frac{\partial \Omega}{\partial \phi^{\bm{I}}} \bigg ] \bigg ] \, , \\
    e^{-\Omega} \, \omega_{\bm{I}} \, e^\Omega &= \sum_{N \geq 0} \frac{(-1)^N}{N!} \bigg [ \underbrace{\Omega \, \ccomma \, \cdots \, \bigg [ \Omega}_{N \text{ copies}} \, \ccomma \; \omega_{\bm{I}} \bigg ] \bigg ] \, ,
\end{align}
the redefinition by $\tau^{(\omega)}$ yields
\begin{equation}
    \omega_{\bm{\bar{p}} \bm{r} \bm{I}} = \Omega_{\bar{\mathtt{p}} \mathtt{r}, \, \bm{I}} + \omega_{\bar{\mathtt{p}} \mathtt{r} \bm{I}} - \left [ \Omega \, \ccomma \, \frac{1}{2} \Omega_{\, , \, \bm{I}} + \omega_{\bm{I}} \right ]_{\bar{\mathtt{p}} \mathtt{r}} + \left [ \Omega \, \ccomma \left [ \Omega \, \ccomma \, \frac{1}{6} \Omega_{\, , \, \bm{I}} + \frac{1}{2} \omega_{\bm{I}} \right ] \right ]_{\bar{\mathtt{p}} \mathtt{r}} - \ldots \, .
\end{equation}
We can arrange on the left-hand side that
\begin{equation}
    \omega_{\bm{\bar{p}} \bm{r} \bm{0}} \vbar_{\mathcal{G}} = 0 \quad \text{and} \quad \omega_{\bm{\bar{p}} \bm{r} (\bm{A_1}, \, \cdots \, \bm{A_N})} \vbar_{\mathcal{G}} = 0 \, ,
\end{equation}
on the line $\mathcal{G} = \{ \phi^{\bm{A}} = 0 \}$ in increasing derivatives of $\phi^{\bm{A}}$, solving for $\Omega_{\bar{\mathtt{p}} \mathtt{r}}^{\vphantom{(1)}} = \Omega_{\bar{\mathtt{p}} \mathtt{r}}^{(1)} + \Omega_{\bar{\mathtt{p}} \mathtt{r}}^{(2)} + \ldots$ on the right-hand side as a series in $\omega_{\bar{\mathtt{p}} \mathtt{r} \bm{I}}$. At first order in the series, we require
\begin{alignat}{2}
    &\Omega_{\bar{\mathtt{p}} \mathtt{r}, \, \bm{0}}^{(1)} \vbar_{\mathcal{G}} &&= - \, \omega_{\bar{\mathtt{p}} \mathtt{r} \bm{0}} \vbar_{\mathcal{G}} \, , \\
    &\Omega_{\bar{\mathtt{p}} \mathtt{r}, \, \bm{A_1} \cdots \, \bm{A_N}}^{(1)} \vbar_{\mathcal{G}} &&= - \, \omega_{\bar{\mathtt{p}} \mathtt{r} (\bm{A_1}, \, \cdots \, \bm{A_N})} \vbar_{\mathcal{G}} \, ,
\end{alignat}
and should therefore set
\begin{equation}
    \Omega_{\bar{\mathtt{p}} \mathtt{r}}^{(1)} = - \int_0^{\phi^{\bm{0}}} d\eta \; \omega_{\bar{\mathtt{p}} \mathtt{r} \bm{0}} \vbar_{\mathcal{G}} - \sum_{N \geq 1} \frac{1}{N!} \, \omega_{\bar{\mathtt{p}} \mathtt{r} (\bm{A_1}, \, \cdots \, \bm{A_N})} \vbar_{\mathcal{G}} \, \phi^{\bm{A_1}} \ldots \, \phi^{\bm{A_N}} \, .
\end{equation}
Here, $\eta$ is an integration variable that runs along $\mathcal{G}$ in place of $\phi^{\bm{0}}$, which instead appears in the limits of integration. At second order, we need
\begin{alignat}{2}
    &\Omega_{\bar{\mathtt{p}} \mathtt{r}, \, \bm{0}}^{(2)} \vbar_{\mathcal{G}} &&= \left [ \Omega^{(1)} \, \ccomma \, \frac{1}{2} \Omega_{\, , \, \bm{0}}^{(1)} + \omega_{\bm{0}}^{\vphantom{(1)}} \right ]_{\bar{\mathtt{p}} \mathtt{r}} \biggvbar_{\mathcal{G}} \, , \\
    &\Omega_{\bar{\mathtt{p}} \mathtt{r}, \, \bm{A_1} \cdots \, \bm{A_N}}^{(2)} \vbar_{\mathcal{G}} &&= \left [ \Omega^{(1)} \, \ccomma \, \frac{1}{2} \Omega_{\, , \, (\bm{A_1}|}^{(1)} + \omega_{(\bm{A_1}|}^{\vphantom{(1)}} \right ]_{\bar{\mathtt{p}} \mathtt{r}, \, |\bm{A_2} \cdots \, \bm{A_N})} \biggvbar_{\mathcal{G}} \, ,
\end{alignat}
and should hence take
\begin{align}
    \Omega_{\bar{\mathtt{p}} \mathtt{r}}^{(2)} &= - \int_0^{\phi^{\bm{0}}} d\eta_1 \; \left [ \int_0^{\eta_1} d\eta_2 \; \omega_{\bm{0}} \vbar_{\mathcal{G}} \, \ccomma \, \frac{1}{2} \omega_{\bm{0}} \vbar_{\mathcal{G}} \right ]_{\bar{\mathtt{p}} \mathtt{r}} \\
    &\quad - \sum_{N \geq 1} \frac{1}{N!} \, \Bigg \{ \left [ \int_0^{\phi^{\bm{0}}} d\eta \; \omega_{\bm{0}} \vbar_{\mathcal{G}} \, \ccomma \, \frac{1}{2} \omega_{(\bm{A_1}, \, \cdots \, \bm{A_N})} \vbar_{\mathcal{G}} \right ]_{\bar{\mathtt{p}} \mathtt{r}} \notag \\
    &\hspace{5.5em} + \sum_{M=1}^{N-1} \binom{N-1}{M} \left [ \omega_{(\bm{A_1}, \, \cdots \, \bm{A_M}|} \vbar_{\mathcal{G}} \, \ccomma \, \frac{1}{2} \omega_{|\bm{A_{M+1}}, \, \cdots \, \bm{A_N})} \vbar_{\mathcal{G}} \right ]_{\bar{\mathtt{p}} \mathtt{r}} \Bigg \} \, \phi^{\bm{A_1}} \ldots \, \phi^{\bm{A_N}} \, . \notag
\end{align}
Iterate and, for a reasonable $\omega_{\bar{\mathtt{p}} \mathtt{r} \bm{I}}$ after the first fermion redefinition, the process will converge near $\mathcal{G}$ to the desired anti-Hermitian $\Omega_{\bar{\mathtt{p}} \mathtt{r}}$. The resulting $\omega_{\bm{\bar{p}} \bm{r} \bm{I}}$ after the second fermion redefinition obeys the last two identities in \cref{eq:k_omega_Fermi_identities}.

Having expended all freedom to redefine the fermion coordinates, the non-vanishing derivatives of $\omega_{\bm{\bar{p}} \bm{r} \bm{I}}$ are now fixed functions of the curvature. We find the relations
\begin{alignat}{2}
    &\omega_{\bm{\bar{p}} \bm{r} \bm{0}, \, \bm{A}} \vbar_{\mathcal{G}} &&= \, - \, R_{\bm{\bar{p}} \bm{r} \bm{0} \bm{A}} \vbar_{\mathcal{G}} \, , \\
    &\omega_{\bm{\bar{p}} \bm{r} \bm{0}, \, \bm{A} \bm{B}} \vbar_{\mathcal{G}} &&= \, - \, R_{\bm{\bar{p}} \bm{r} \bm{0} (\bm{A}; \, \bm{B})} \vbar_{\mathcal{G}} \, , \\
    & &&\hspace{0.52em} \vdots \notag
\end{alignat}
which are the analogs on $\mathcal{E}$ of \crefrange{eq:Fermi_identities_2}{eq:Fermi_identities_3} on $T\mathcal{M}$.\footnote{To make the analogy explicit, invoke the additional identity $R_{\bm{\bar{p}} \bm{r} \bm{I} \bm{J}} = - 2 R_{\bm{\bar{p}} \bm{I} \bm{J} \bm{r}} = - 2 R_{\bm{\bar{p}} \bm{J} \bm{r} \bm{I}}$ up to $\mathcal{O}(\chi^2)$ for the graded metric $G$. There is no cost to treating the two-fermion curvature components, as well as their covariant derivatives, as those of $G$.} Like the scalar sector, the identities on the vector bundle connection hold on the whole of $\mathcal{G}$ and can be differentiated to arbitrary order in $\phi^{\bm{0}}$.

\subsection{Probing field space with scattering amplitudes}

Having established a desirable coordinate system on field space, we move on to the detailed calculation of field redefinition invariant quantities, revealing how the geometric information along a geodesic enters into physical observables. On-shell scattering amplitudes are our primary observable of interest, and we compute them using Feynman rules expressed in Fermi normal coordinates. For the Lagrangian \cref{eq:Lagrangian}, they read:
\begin{align}
    \vcenter{\hbox{ \begin{tikzpicture} \begin{feynman}
            \vertex (I) at (-0.8,0) {$\bm{I}$};
            \vertex (J) at (0.8,0) {$\bm{J}$};
            \diagram* {(I) -- [scalar, edge label=\(q\)] (J)};
        \end{feynman} \end{tikzpicture}}}
    &\; = i \, g^{\bm{I} \bm{J}} \vbar \; \frac{1}{q^2 - m_{\bm{I}}^2} \, , \label{eq:scalar_propagator} \\
    \vcenter{\hbox{ \begin{tikzpicture} \begin{feynman}
            \vertex (r) at (-0.8,0) {$\bm{r}$};
            \vertex (p) at (0.8,0) {$\bm{\bar{p}}$};
            \diagram* {(r) -- [edge label=\(q\)] (p)};
        \end{feynman} \end{tikzpicture}}}
    &\; = i \, k^{\bm{r} \bm{\bar{p}}} \vbar \; \frac{\slashed{q} + m_{\bm{r}}}{q^2 - m_{\bm{r}}^2} \, , \\
    \vcenter{ \hbox{ \begin{tikzpicture} \begin{feynman}
        \vertex (m) at (0.00, 0.00);
        \vertex (I1) at (-0.67, 1.00) {$\bm{I_1}$};
        \vertex (I2) at (-1.20, -0.00) {$\bm{I_2}$};
        \vertex (In) at (0.46, 1.11) {$\bm{I_N}$};
        \diagram* {
        (I1) -- [scalar] (m) -- [scalar] (In),
        (I2) -- [scalar] (m)};
        \draw [dotted, thick, domain=215:395] plot ({0.6*cos(\x)}, {0.6*sin(\x)});
    \end{feynman} \end{tikzpicture}}}
    &\; = - \, i \, \sum_{J < K}^N g_{\bm{I_J} \bm{I_K} \, , \, \bm{I_1} \, \cdots \, \widehat{\bm{I_J} \bm{I_K}} \, \cdots \, \bm{I_N}} \vbar \; (q_{I_J} \cdot q_{I_K}) - i \, V_{ \, , \, \bm{I_1} \, \cdots \, \bm{I_N}} \vbar \, , \\
    \vcenter{ \hbox{ \begin{tikzpicture} \begin{feynman}
        \vertex (m) at (0.00, 0.00);
        \vertex (p) at (-0.67, 1.00) {$\bm{\bar{p}}$};
        \vertex (r) at (-1.20, -0.00) {$\bm{r}$};
        \vertex (I1) at (-0.67, -1.00) {$\bm{I_1}$};
        \vertex (I2) at (0.46, -1.11) {$\bm{I_2}$};
        \vertex (In) at (0.46, 1.11) {$\bm{I_N}$};
        \diagram* {
        (p) -- (m) -- (r),
        (I1) -- [scalar] (m) -- [scalar] (In),
        (I2) -- [scalar] (m)};
        \draw [dotted, thick, domain=325:395] plot ({0.6*cos(\x)}, {0.6*sin(\x)});
    \end{feynman} \end{tikzpicture}}}
    &\; = \frac{i}{2} \, k_{\bm{\bar{p}} \bm{r}, \, \bm{I_1} \cdots \bm{I_N}} \vbar \; ( \slashed{q}_r - \slashed{q}_{\bar{p}} ) + i \, \sum_J^N \omega_{\bm{\bar{p}} \bm{r} \bm{I_J} \, , \, \bm{I_1} \, \cdots \, \widehat{\bm{I_J}} \, \cdots \, \bm{I_N}} \vbar \; \slashed{q}_{I_J} - i \, M_{\bm{\bar{p}} \bm{r}, \, \bm{I_1} \, \cdots \, \bm{I_N}} \vbar \, . \label{eq:fermion_vertex}
\end{align}
All momenta are ingoing, hats indicate omitted scalar indices and vertical bars denote evaluation at the origin. The non-tensorial partial derivatives on field space that appear can be converted into covariant ones through repeated application of the identities from the previous section. The Feynman rules are then manifestly geometric, and so are the amplitudes assembled from them.

The explicit conversion formulas are as follows. Substituting the scalar kinetic term and potential into \cref{eq:cov_deriv} and covariantizing any partial derivative of the connection using \crefrange{eq:Fermi_identities_0}{eq:Fermi_identities_3}, we obtain
\begingroup
\allowdisplaybreaks
\begin{alignat}{3}
    &&\partial_{\bm{A}}^0: \quad &g_{\bm{I} \bm{J}, \, \bm{0}^{N+1}} \vbar &&= \, 0 \, , \\
    &&\partial_{\bm{A}}^1: \quad &g_{\bm{I} \bm{J}, \, \bm{A} \bm{0}^N} \vbar &&= \, 0 \, , \\
    &&\partial_{\bm{A}}^2: \quad &g_{\bm{0} \bm{0}, \, \bm{A} \bm{B} \bm{0}^N} \vbar &&= \, 2 \, R_{\bm{0} (\bm{A} \bm{B}) \bm{0}; \, \bm{0}^N} \vbar \, , \\
    & &&g_{\bm{A} \bm{0}, \, \bm{B} \bm{C} \bm{0}^N} \vbar &&= \, \frac{4}{3} \, R_{\bm{A} (\bm{B} \bm{C}) \bm{0}; \, \bm{0}^N} \vbar \, , \\
    & &&g_{\bm{A} \bm{B}, \, \bm{C} \bm{D} \bm{0}^N} \vbar &&= \, \frac{2}{3} \, R_{\bm{A} (\bm{C} \bm{D}) \bm{B}; \, \bm{0}^N} \vbar \, , \\
    &&\partial_{\bm{A}}^3: \quad &g_{\bm{0} \bm{0}, \, \bm{A} \bm{B} \bm{C} \bm{0}^N} \vbar &&= \, 2 \, R_{\bm{0} (\bm{A} \bm{B}| \bm{0}; \, |\bm{C}) \bm{0}^N} \vbar \, , \\
    & &&g_{\bm{A} \bm{0}, \, \bm{B} \bm{C} \bm{D} \bm{0}^N} \vbar &&= \, \frac{3}{2} \, R_{\bm{A} (\bm{B} \bm{C}| \bm{0}; \, |\bm{D}) \bm{0}^N} \vbar \, , \\
    &&\partial_{\bm{A}}^4: \quad &g_{\bm{0} \bm{0}, \, \bm{A} \bm{B} \bm{C} \bm{D} \bm{0}^N} \vbar &&= \, 2 \, R_{\bm{0} (\bm{A} \bm{B}| \bm{0}; \, |\bm{C} \bm{D}) \bm{0}^N} \vbar + 8 \left ( \tensor{R}{_{\bm{K} (\bm{A} \bm{B}| \bm{0}}} \, \tensor{R}{^{\bm{K}}_{|\bm{C} \bm{D}) \bm{0}}} \, \right )_{; \, \bm{0}^N} \biggvbar \, , \\
    & && &&\hspace{0.52em} \vdots \notag
\end{alignat}%
\endgroup
and
\begin{equation}
    V_{, \, \bm{A_1} \cdots \bm{A_M} \bm{0}^N} \vbar = \, V_{; \, (\bm{A_1} \cdots \bm{A_M}) \bm{0}^N} \vbar \, .
\end{equation}
for all $M, N \geq 0$. Note that scalar vertices with no more than one $\phi^{\bm{A}}$ leg are momentum-independent in these coordinates. The same procedure on the two-fermion curvature components and the fermion mass matrix yields
\begin{alignat}{3}
    &&\partial_{\bm{A}}^0: \quad &\omega_{\bm{\bar{p}} \bm{r} \bm{I} \, , \, \bm{0}^N} \vbar &&= \, 0 \, , \\
    &&\partial_{\bm{A}}^1: \quad &\omega_{\bm{\bar{p}} \bm{r} \bm{0} \, , \, \bm{A} \bm{0}^N} \vbar &&= \, - \, R_{\bm{\bar{p}} \bm{r} \bm{0} \bm{A}; \, \bm{0}^N} \vbar \, , \\
    & &&\omega_{\bm{\bar{p}} \bm{r} \bm{A} \, , \, \bm{B} \bm{0}^N} \vbar &&= \, - \, \frac{1}{2} \, R_{\bm{\bar{p}} \bm{r} \bm{A} \bm{B}; \, \bm{0}^N} \vbar \, , \\
    &&\partial_{\bm{A}}^2: \quad &\omega_{\bm{\bar{p}} \bm{r} \bm{0} \, , \, \bm{A} \bm{B} \bm{0}^N} \vbar &&= \, - \, R_{\bm{\bar{p}} \bm{r} \bm{0} (\bm{A}; \, \bm{B}) \bm{0}^N} \vbar \, , \\
    & && &&\hspace{0.52em} \vdots \notag
\end{alignat}
and
\begin{equation}
    M_{\bm{\bar{p}} \bm{r}, \, \bm{A_1} \cdots \bm{A_M} \bm{0}^N} \vbar = \, M_{\bm{\bar{p}} \bm{r}; \, (\bm{A_1} \cdots \bm{A_M}) \bm{0}^N} \vbar \, .
\end{equation}
There is no contribution from the trivialized $k_{\bm{\bar{p}} \bm{r}}$ to the fermion vertices.

We give some examples of tree-level amplitudes assembled in this manner. With the goal of probing various curvature components on $\mathcal{M}$ and $\mathcal{E}$, we choose the following initial and final states:
\begin{alignat}{3}
    && \text{scalars:} &\quad \mathcal{A} \, \Big ( \phi_{\bm{A}}^{\vphantom{N+2}} \, \phi_{\bm{A}}^{\vphantom{N+2}} \rightarrow \phi_{\bm{0}}^{N+2} \Big ) \, , &&\mathcal{A} \, \Big ( \phi_{\bm{A}}^{\vphantom{N}} \, \phi_{\bm{A}}^{\vphantom{N}} \rightarrow \phi_{\bm{B}}^{\vphantom{N}} \, \phi_{\bm{B}}^{\vphantom{N}} \, \phi_{\bm{0}}^N \Big ) \, , \\
    && \text{fermions:} &\quad \mathcal{A} \, \Big ( \overbar{\psi}_{\bm{\bar{p}}}^{\vphantom{N+1}} \, \psi_{\bm{r}}^{\vphantom{N+1}} \rightarrow \phi_{\bm{A}}^{\vphantom{N+1}} \, \phi_{\bm{0}}^{N+1} \Big ) \, , \quad &&\mathcal{A} \, \Big ( \overbar{\psi}_{\bm{\bar{p}}}^{\vphantom{N}} \, \psi_{\bm{r}}^{\vphantom{N}} \rightarrow \phi_{\bm{A}}^{\vphantom{N}} \, \phi_{\bm{B}}^{\vphantom{N}} \, \phi_{\bm{0}}^N \Big ) \, ,
\end{alignat}
and examine their large-momentum limit, as determined by the kinetic terms of the theory. Since different external particles have the same flavor, let us distinguish their momenta by labels from $1$ to $N+4$ for the initial states and the final $\phi_{\bm{A}}$ and $\phi_{\bm{0}}$ particles in that order, \emph{i.e.} as written above from left to right. The leading terms in momentum of the first scalar amplitude are
\begin{equation}
    \mathcal{A} \, \Big ( \phi_{\bm{A}}^{\vphantom{N+2}} \, \phi_{\bm{A}}^{\vphantom{N+2}} \rightarrow \phi_{\bm{0}}^{N+2} \Big ) \supset - \, \tensor{R}{_{\bm{A} \bm{0} \bm{0} \bm{A}; \, \bm{0}^N}} \vbar \, s_{12} - \tensor{R}{_{\bm{C} \bm{0} \bm{0} \bm{A} \vphantom{\bm{0}^{N-2}}}} \, \tensor{R}{^{\bm{C}}_{\bm{0} \bm{0} \bm{A}; \, \bm{0}^{N-2}}} \vbar \, \frac{s_{34} \, s_{1234}}{s_{134}} + \ldots \, , \label{eq:A_pi2}
\end{equation}
where $s_{c \cdots d} = (q_c + \ldots + q_d)^2$ are Mandelstam variables. The first term originates from the contact diagram, while the rest come from $t$-channel diagrams with $\phi_{\bm{C}}$ propagators and at least two $\phi_{\bm{0}}$ legs at each vertex.\footnote{We classify contact and propagator terms in an amplitude according to the Feynman rule prescriptions in \crefrange{eq:scalar_propagator}{eq:fermion_vertex}.} The leading pieces of the second scalar amplitude read
\begin{align}
    &\mathcal{A} \, \Big ( \phi_{\bm{A}}^{\vphantom{N}} \, \phi_{\bm{A}}^{\vphantom{N}} \rightarrow \phi_{\bm{B}}^{\vphantom{N}} \, \phi_{\bm{B}}^{\vphantom{N}} \, \phi_{\bm{0}}^N \Big ) \supset - \frac{1}{2} \, R_{\bm{A} \bm{B} \bm{B} \bm{A}; \, \bm{0}^N} \vbar \left ( s_{12} - \frac{1}{3} s_{1234} + s_{34} \right ) \label{eq:A_pi4} \\
    &\hspace{5em} - R_{\bm{A} \bm{B} \bm{0} (\bm{A}; \, \bm{B}) \bm{0}^{N-1}} \vbar \left ( s_{12} - s_{34} \right ) \notag \\
    &\hspace{5em} - \left [ \tensor{R}{^{\vphantom{\bm{B}}}_{(\bm{A}| \bm{0} \bm{0}| \bm{A}; \, \bm{B} \bm{B}) \bm{0}^{N-2}}} + 4 \left ( \tensor{R}{_{\bm{K} (\bm{A} \bm{A}| \bm{0}}} \, \tensor{R}{^{\bm{K}}_{|\bm{B} \bm{B}) \bm{0}}} \, \right )_{; \, \bm{0}^{N-2}} \right ] \biggvbar \; s_{1234} \notag \\
    &\hspace{5em} - \frac{1}{3} \, \tensor{R}{^{\vphantom{\bm{B}}}_{\bm{K} \bm{B} \bm{B} \bm{A}}} \, \tensor{R}{^{\bm{K}}_{\bm{0} \bm{0} \bm{A}; \, \bm{0}^{N-4}}} \vbar \, \frac{(2 s_{34} - s_{13} - s_{14}) \, s_{1234}}{s_{134}} + \ldots \, . \notag
\end{align}
The contact diagram yields the first three lines. While we have chosen to compute the amplitudes in Fermi normal coordinates, the resulting expressions necessarily equal those in Riemann normal coordinates found in \emph{e.g.} \cite{Cohen:2021ucp}; the use of a different parameterization here shuffles contributions between Feynman diagrams and eliminates contact terms that are quadratic and above in curvature to a greater extent. Similarly, we obtain the leading pieces
\begin{align}
    &\hspace{0.55em} \mathcal{A} \, \Big ( \overbar{\psi}_{\bm{\bar{p}}}^{\vphantom{N+1}} \, \psi_{\bm{r}}^{\vphantom{N+1}} \rightarrow \phi_{\bm{A}}^{\vphantom{N+1}} \, \phi_{\bm{0}}^{N+1} \Big ) \supset R_{\bm{\bar{p}} \bm{r} \bm{0} \bm{A}; \, \bm{0}^N} \vbar \, \overbar{v}(q_1) \, \slashed{q}_3 \, u(q_2) \label{eq:A_ttpi} \\
    &\hspace{5em} - \tensor{R}{^{\vphantom{\bm{C}}}_{\bm{\bar{p}} \bm{r} \bm{0} \bm{C} \vphantom{\bm{0}^{N-2}}}} \, \tensor{R}{^{\bm{C}}_{\bm{0} \bm{0} \bm{A}; \, \bm{0}^{N-2}}} \vbar \, \overbar{v}(q_1) \, \slashed{q}_4 \, \frac{s_{1234}}{s_{124}} \, u(q_2) + \ldots \, , \notag \\
    &\mathcal{A} \, \Big ( \overbar{\psi}_{\bm{\bar{p}}}^{\vphantom{N}} \, \psi_{\bm{r}}^{\vphantom{N}} \rightarrow \phi_{\bm{A}}^{\vphantom{N}} \, \phi_{\bm{B}}^{\vphantom{N}} \, \phi_{\bm{0}}^N \Big ) \supset \frac{1}{2} \, R_{\bm{\bar{p}} \bm{r} \bm{A} \bm{B}; \, \bm{0}^N} \vbar \, \overbar{v}(q_1) \, ( - \slashed{q}_3 + \slashed{q}_4 ) \, u(q_2) \label{eq:A_ttpi2} \\
    &\hspace{5em} + R_{\bm{\bar{p}} \bm{r} \bm{0} (\bm{A}; \, \bm{B}) \bm{0}^{N-1}} \vbar \, \overbar{v}(q_1) \, ( \slashed{q}_3 + \slashed{q}_4 ) \, u(q_2) \notag \\
    &\hspace{5em} + \tensor{R}{^{\vphantom{\bm{s}}}_{\bm{\bar{p}} \bm{s} \bm{0} \bm{A} \vphantom{\bm{0}^{N-2}}}} \, \tensor{R}{^{\bm{s}}_{\bm{r} \bm{0} \bm{B}; \, \bm{0}^{N-2}}} \vbar \, \overbar{v}(q_1) \, \frac{\slashed{q}_5 \, (\slashed{q}_1 + \slashed{q}_3 + \slashed{q}_5) \, (\slashed{q}_1 + \slashed{q}_3 + \slashed{q}_4 + \slashed{q}_5)}{s_{135}} \, u(q_2) + \ldots \, , \notag
\end{align}
in the fermion amplitudes, where $\overbar{v}$ and $u$ are Dirac spinor wavefunctions.\footnote{Not to be confused with $v$ for a real-valued vacuum expectation value.}

We now see how the the geometry of field space is reflected in scattering processes --- for a given curvature component, the value of its covariant derivative at the vacuum is encoded in a contact term of the amplitude between states of the corresponding flavors. There may also be curvature terms involving different index permutations or contractions between lower-order covariant derivatives, as well as non-curvature terms arising from the scalar potential and the fermion mass matrix that are subleading in momentum. These geometric pieces prepend different kinematic factors and can be extracted from the differential cross section in different regions of phase space, allowing us to decipher the curvature at the vacuum.

Going beyond the vacuum, an important qualitative geometric feature besides the precise values of the curvature is the presence of singularities, a hallmark of states that have been integrated out from the EFT \cite{Falkowski:2019tft, Cohen:2020xca, Cohen:2021ucp}. This is where the advantages of Fermi normal coordinates, adapted to a geodesic $\mathcal{G}$ originating from the vacuum, become most apparent. Its tangent vector basis $\{ \partial_{\bm{I}}, \partial_{\bm{r}}, \partial_{\bm{\bar{p}}} \}$ forms a covariantly constant orthonormal frame along $\mathcal{G}$, so that the components of $R$ in this basis reveal any parallelly propagated curvature singularity that $\mathcal{G}$ eventually runs into \cite{Clarke:1973local, Hawking:1973uf, Ellis:1974ug, Ellis:1977pj, Clarke:1982local}. Moreover, should there indeed be such a distant singularity, the flatness of this coordinate system along $\mathcal{G}$ makes it easy to detect at the starting point without hindrance from the curvature in between. Using the vector bundle curvature
\begin{equation}
    R_{\bm{\bar{p}} \bm{r} \bm{0} \bm{A}} \vbar_{\mathcal{G}} = \sum_{N \geq 0} \frac{1}{N!} \, R_{\bm{\bar{p}} \bm{r} \bm{0} \bm{A}, \, \bm{0}^N} \vbar \, \phi_{\bm{0}}^N \, ,
\end{equation}
as an example, the Cauchy-Hadamard theorem on the complex plane $\phi_{\bm{0}} \in \mathbb{C}$ states that a singularity at proper distance $\phi_{\bm{0}}^* \in \mathbb{R}_{\geq 0}$ requires the Taylor coefficients at the vacuum $\phi_{\bm{0}} = 0$ to grow as
\begin{equation}
    \frac{1}{\phi_{\bm{0}}^*} = \limsup_{N \rightarrow \infty} \, \left | \frac{1}{N!} \, R_{\bm{\bar{p}} \bm{r} \bm{0} \bm{A}, \, \bm{0}^N} \vbar \, \right |^{\frac{1}{N}} \, .
\end{equation}
Fermi normal coordinates equate the partial $\phi_{\bm{0}}$ derivatives with covariant ones that appear in field redefinition invariant amplitudes, \emph{i.e.}
\begin{equation}
    R_{\bm{\bar{p}} \bm{r} \bm{0} \bm{A}, \, \bm{0}^N} \vbar = R_{\bm{\bar{p}} \bm{r} \bm{0} \bm{A}; \, \bm{0}^N} \vbar \subset \mathcal{A} \, \Big ( \overbar{\psi}_{\bm{\bar{p}}}^{\vphantom{N+1}} \, \psi_{\bm{r}}^{\vphantom{N+1}} \rightarrow \phi_{\bm{A}}^{\vphantom{N+1}} \, \phi_{\bm{0}}^{N+1} \Big ) \, . \label{eq:fermi_curv_in_amp}
\end{equation}
One can thereby identify singularities along $\mathcal{G}$ by measuring scattering amplitudes with an increasing number of $\phi_{\bm{0}}$ states.

Alternative parameterizations like Riemann normal coordinates complicate the inference of geometric singularities from scattering amplitudes in two ways. At the first equality sign in \cref{eq:fermi_curv_in_amp}, they generate quadratic and higher-order terms from the application of \cref{eq:cov_deriv} due to curvature along $\mathcal{G}$.\footnote{These terms were assembled in \cite{Cohen:2021ucp} into HEFT form factors via a recursion relation. From the perspective of the current work, the recursion relation reflects the equivalence of Fermi partial and covariant derivatives as seen from another coordinate system.} At the second inclusion sign, there may also be additional higher-order contact terms in curvature that appear at six points and above.\footnote{The presence of such terms cannot be entirely avoided, evident from the third line of \cref{eq:A_pi4}.} There are hence more terms to tease apart within the scattering amplitude to extract the Taylor coefficients we need. While on-shell amplitudes are ultimately independent of field parameterization, Fermi normal coordinates represent a rewriting that makes the connection to field space geometry most transparent.

The deduction of singularities from the infinite multiplicity limit is most conservative, and most impractical. In taking $N \rightarrow \infty$, several issues arise from using scattering amplitudes as a proxy to curvature:
\begin{itemize}
    \item This formal limit is strictly impossible to attain if the $\phi_{\bm{0}}$ state is massive.
    \item There is a combinatorial number of propagator diagrams of the same order in momentum that compete with the Taylor coefficient we want to extract.
    \item There is also a combinatorial number of terms subleading in momentum that are in competition. The energy of the scattering process must be large enough to suppress them, and higher-derivative operators beyond those captured by the truncated Lagrangian \cref{eq:Lagrangian} may then become important.
\end{itemize}
A reduction to more realistic multiplicities is possible if we restrict our attention to pole-like singularities
\begin{equation}
    \frac{1}{(\phi_{\bm{0}}^* - \phi_{\bm{0}}^{\vphantom{*}})^d} = \sum_{N \geq 0} \frac{(-1)^N}{(\phi_{\bm{0}}^*)^{N+d}} \, \binom{-d \,}{N} \, \phi_{\bm{0}}^N \, , \label{eq:order-d-pole}
\end{equation}
for some small $d$, which is nevertheless generic to EFTs arising from heavy states that are integrated out at tree level.\footnote{The expansion holds for arbitrary $d$ if we use generalized binomial coefficients.} The rigorous result
\begin{equation}
    \frac{1}{\phi_{\bm{0}}^*} = \limsup_{N \rightarrow \infty} \, \left | \frac{1}{(\phi_{\bm{0}}^*)^{N+d}} \, \binom{-d \,}{N} \right |^{\frac{1}{N}} \, ,
\end{equation}
can then be parametrically approximated if we take $N \gtrsim d$. This corroborates with constraints on scattering amplitudes due to perturbative unitarity, which applies regardless of multiplicity. Barring some predictable scaling with $N$, the unitarity violation scales $E_N^*$ across $N$ are all estimates of the same heavy state mass. Since the energy scaling of scattering amplitudes is controlled by the Taylor coefficients of the curvature, the size of a lower-order one can plausibly be indicative of higher orders.

The relation between singularity and unitarity can be made precise as follows. Consider the unitarity bound $\sigma_N \leq 4 \pi / E^2$ where $\sigma_N$ is the $N+4$ particle total cross section and $E = \sqrt{s_{12}}$ is the center of mass energy \cite{Maltoni:2001dc, Dicus:2004rg}. The cross section can be obtained from the scattering amplitude by integrating over the phase space of the $N+2$ final states. Based on a closed form expression for massless final states, which is derived in \cref{app:phase_space}, we have the general formula
\begin{align}
    \int d\mathrm{LIPS} \; q_{\bm{I_J}}^\mu \, q_{\bm{I_K}}^{\vphantom{\mu} \nu} = \rho_{N+2} \times \alpha_{N+2} &\times \frac{1}{8 \pi \, N! \, (N+3)!} \left ( \frac{E}{4\pi} \right )^{2N} \label{eq:phase_space_integral} \\
    &\times \begin{cases}
        2 \left ( Q^\mu \, Q^\nu - \frac{1}{4} \, E^2 \, \eta^{\mu\nu} \right ) &\text{if } \bm{J} = \bm{K} \\
        \left ( Q^\mu \, Q^\nu + \frac{1}{2(N+1)} \, E^2 \, \eta^{\mu\nu} \right ) &\text{otherwise}
    \end{cases} \, , \notag
\end{align}
where $Q^\mu$ is the total momentum and $\eta^{\mu\nu}$ is the Minkowski metric. The symmetry factor $\rho_{N+2}$ accounts for any identical final states, while the dimensionless factor $\alpha_{N+2}$ accounts for the true masses of the final states, ranging from $1$ when they are all massless to $0$ when they saturate $E$. For the example process above with fermions of definite helicity, in the case that the contact term dominates, we find the cross section
\begin{equation}
    \sigma_N \sim \alpha_{N+2} \, \frac{\pi}{N! \, (N+1)! \, (N+3)!} \left ( R_{\bm{\bar{p}} \bm{r} \bm{0} \bm{A}; \, \bm{0}^N} \vbar \, \right )^2 \left ( \frac{E}{4 \pi} \right)^{2N+2} + \mathcal{O}(E^{2N}) \, ,
\end{equation}
the scale of unitarity violation
\begin{equation}
    E_N^* \sim 4 \pi \left ( \frac{N}{e} \right )^{\frac{3}{2}} \, \Big | R_{\bm{\bar{p}} \bm{r} \bm{0} \bm{A}; \, \bm{0}^N} \vbar \, \Big |^{-\frac{1}{N+2}} \, ,
\end{equation}
and the proper distance of the curvature singularity
\begin{equation}
    \phi_{\bm{0}}^* \sim \frac{\sqrt{e}}{4\pi} \, \frac{E_N^*}{\sqrt{N}} \, . \label{eq:singularity_from_unitarity}
\end{equation}
Indeed, $\alpha_{N+2}$ should be close to unity for moderate values of $N$, since the heavy state should be substantially more massive than the remaining states in the EFT. Therefore, we can reasonably estimate the position of a singularity in the above curvature using the scale of unitarity violation at finite multiplicity.\footnote{We also learn that $E_N^* \sim \sqrt{N}$ in this range of $N$. This scaling necessarily breaks down at very large $N$, when $\alpha_{N+2}$ deviates significantly from unity.} Competing contributions to the cross section make the full analysis more complicated; we will return to this issue in the next section and see that specializing to the geometry of the Higgs sector can buy us some simplifications.

\section{Field space geometry of the Higgs sector}
\label{sec:application}

With a refined framework for field redefinition invariance, a distinguished computational basis, and a practical probe of geometry in hand, we turn to what is arguably their most important phenomenological application. Motivated by the SM Higgs sector and the importance of deciphering the physical content of its EFT description, we translate the EFT into a geometric language and examine the encoded information associated with the various degrees of freedom, namely the physical Higgs and Goldstone bosons together with the SM fermions. Lifting existing restrictions in the literature, we impose only symmetries mandated by the electroweak gauge group $SU(2)_L \times U(1)_Y$, which introduce new wrinkles to the base space geometry. We remain agnostic on whether the symmetry is realized linearly, and must therefore enumerate all HEFT operators that fit into the Lagrangian \cref{eq:Lagrangian}. This is mostly easy done in the well-known Callan-Coleman-Wess-Zumino (CCWZ) coordinates on field space \cite{Callan:1969sn}.

Let us delineate the particle content, paying close attention to their transformation properties under electroweak symmetry. Denoting Pauli matrices as $\sigma_A$, we conveniently rewrite the $SU(2)_L$ Higgs doublet $H$ as a $2 \times 2$ matrix $\Sigma = ( i \sigma_2 H^* , H)$. This matrix transforms as $\Sigma \rightarrow L \Sigma \overbar{R}$ with $L \in SU(2)_L$ and $R \in U(1)_Y$, the latter being a diagonal subgroup of an approximate $SU(2)_R$ symmetry. The vacuum lies at $\langle \Sigma \rangle = v / \sqrt{2}$ and serves as the origin about which we expand
\begin{equation}
    \Sigma = \frac{v+h}{\sqrt{2}} \, U \quad \text{where} \quad U = \exp \left [ i \left ( \sin^{-1} \frac{|\vec{\pi}|}{v} \right ) \, \hat{\pi} \cdot \vec{\sigma} \right ] \, ,
\end{equation}
via a polar decomposition.\footnote{This particular parameterization of $U$ is chosen in view of the conventions of \cite{Cohen:2021ucp}.} The field space coordinates here are the physical Higgs boson $h$ and the Goldstone bosons $\vec{\pi} = (\pi_1, \pi_2, \pi_3)$; the latter parameterize a unitary matrix $U$ that transforms as $U \rightarrow L U \overbar{R}$, and the matrix exponential makes the symmetry non-linear. We can additionally form two traceless, Hermitian matrices
\begin{equation}
    T = U \sigma_3 \overbar{U} \quad \text{and} \quad V_\mu = i (\partial_\mu U) \overbar{U} \, .
\end{equation}
Both have definite transformations $T \rightarrow L T \overbar{L}$ and $V_\mu \rightarrow L V_\mu \overbar{L}$, but $T$ manifestly violates $SU(2)_R$ symmetry. Our inclusion of $T$ into the theory extends previous work that assumes custodial symmetry.

As for the SM fermions, we group them into $SU(2)_L$ and $SU(2)_R$ doublets
\begin{equation}
    Q_L = \begin{pmatrix}
        t_L \\ b_L
    \end{pmatrix}
    \quad \text{and} \quad
    Q_R = \begin{pmatrix}
        t_R \\ b_R
    \end{pmatrix} \, ,
\end{equation}
which transform as $Q_L \rightarrow L Q_L$ and $Q_R \rightarrow R Q_R$. One generation of quarks, \emph{e.g.} top and bottom, is sufficient to elicit the bundle structure of fermionic field space, which will be our primary interest.\footnote{There is no inherent difficulty in incorporating leptons or multiple generations of fermions; these manifest as the vector space structure within each fiber of the bundle.} The four $t$ and $b$ doublet components, together with their four conjugates, constitute the fermionic field space coordinates. Using the building blocks $h$, $U$, $T$, $V_\mu$, $Q_L$ and $Q_R$, we can now assemble operators invariant under electroweak symmetry systematically \cite{Feruglio:1992wf, Appelquist:1993ka, Bagger:1993zf, Koulovassilopoulos:1993pw, Alonso:2012px, Buchalla:2013rka, Brivio:2013pma}. The singlet $h$ endures no restrictions, while the other building blocks must appear in specific combinations.

\subsection{Scalar sector}

The invariant operators in the scalar sector up to two derivatives read\footnote{With geometry in mind, we have chosen conventions for the $h$-dependent functions that differ mildly from literature, rendering $K$ and the $F$'s dimensionless in particular. We may subsequently leave their $h$-dependence implicit or evaluate them at $h=0$, as should be clear from context.}
\begin{subequations}
\begin{align}
    \mathcal{L} &\supset - \, V(h) + \frac{1}{2} \, K(h) \; \partial_\mu h \, \partial^\mu h + \frac{1}{4} \, v^2 \, F_C(h) \Tr \left [ V_\mu V^\mu \right ] \\
    &\quad + \frac{1}{8} \, v^2 \, F_T(h) \Tr \left [ V_\mu T \right ] \Tr \left [ V^\mu T \right ] + \frac{1}{2} \, v \, F_{2D}(h) \Tr \left [ V_\mu T \right ] \, \partial^\mu h \, .
\end{align}
\end{subequations}
Those on the second line violate custodial symmetry and the last is CP-odd; they are expected to be small near the vacuum based on experimental bounds. Reading off the metric on $\mathcal{M}$, we find a block form
\begin{equation}
    g_{IJ} = 
    \begingroup
    \setlength\arraycolsep{6pt}
    \renewcommand\arraystretch{1.4}
    \begin{pmatrix}
        g_{hh}(K) & g_{h \pi_B}(F_{2D}) \\
        g_{\pi_A h}(F_{2D}) & g_{\pi_A \pi_B}(F_C, F_T) \\
    \end{pmatrix}
    \endgroup
    \, .
\end{equation}
The scalar field space is locally a product space $\mathbb{R} \times S^3$ charted by the $h$ and $\pi_A$ coordinates respectively; $K$ and $F_C$ scale the $h$ and $\pi_A$ directions respectively, $F_T$ deforms the $\pi_3$ direction relative to the other Goldstone bosons, and $F_{2D}$ couples $h$ to all $\pi_A$ but mostly $\pi_3$. Full expressions for the metric components read
\begingroup
\allowdisplaybreaks
\begin{alignat}{3}
    &g_{hh} &&= \, K \, , && \\
    &g_{h \pi_1} && = && \hspace{1.43em} \frac{F_{2D}}{v} \left ( \pi_2 - \frac{\pi_1 \pi_3}{\sqrt{v^2 - \vec{\pi}^2}} \right ) \, , \\
    &g_{h \pi_2} && = && \, - \frac{F_{2D}}{v} \left ( \pi_1 + \frac{\pi_2 \pi_3}{\sqrt{v^2 - \vec{\pi}^2}} \right ) \, , \\
    &g_{h \pi_3} && = && \, - \hspace{0.12em} F_{2D} \hspace{0.12em} \left ( 1 + \frac{\pi_3^2}{v^2 - \vec{\pi}^2} \right ) \sqrt{1 - \frac{\vec{\pi}^2}{v^2}} \, , \\
    &g_{\pi_1 \pi_1} &&= \, F_C \left ( 1 + \frac{\pi_1^2}{v^2 - \vec{\pi}^2} \right ) && \, + \frac{F_T}{v^2} \enspace \left ( \pi_2 - \frac{\pi_1 \pi_3}{\sqrt{v^2 - \vec{\pi}^2}} \right )^2 \, , \\
    &g_{\pi_1 \pi_2} &&= \, F_C \hspace{0.4em} \frac{\pi_1 \pi_2}{v^2 - \vec{\pi}^2} && \, - \frac{F_T}{v^2} \enspace \left ( \pi_1 + \frac{\pi_2 \pi_3}{\sqrt{v^2 - \vec{\pi}^2}} \right ) \left ( \pi_2 - \frac{\pi_1 \pi_3}{\sqrt{v^2 - \vec{\pi}^2}} \right ) \, , \\
    &g_{\pi_1 \pi_3} &&= \, F_C \hspace{0.4em} \frac{\pi_1 \pi_3}{v^2 - \vec{\pi}^2} && \, - \frac{F_T}{v} \enspace \left ( 1 + \frac{\pi_3^2}{v^2 - \vec{\pi}^2} \right ) \sqrt{1 - \frac{\vec{\pi}^2}{v^2}} \left ( \pi_2 - \frac{\pi_1 \pi_3}{\sqrt{v^2 - \vec{\pi}^2}} \right ) \, , \\
    &g_{\pi_2 \pi_2} &&= \, F_C \left ( 1 + \frac{\pi_2^2}{v^2 - \vec{\pi}^2} \right ) && \, + \frac{F_T}{v^2} \enspace \left ( \pi_1 + \frac{\pi_2 \pi_3}{\sqrt{v^2 - \vec{\pi}^2}} \right )^2 \, , \\
    &g_{\pi_2 \pi_3} &&= \, F_C \hspace{0.4em} \frac{\pi_2 \pi_3}{v^2 - \vec{\pi}^2} \, && \, + \frac{F_T}{v} \enspace \left ( 1 + \frac{\pi_3^2}{v^2 - \vec{\pi}^2} \right ) \sqrt{1 - \frac{\vec{\pi}^2}{v^2}} \left ( \pi_1 + \frac{\pi_2 \pi_3}{\sqrt{v^2 - \vec{\pi}^2}} \right ) \, , \\
    &g_{\pi_3 \pi_3} &&= \, F_C \left ( 1 + \frac{\pi_3^2}{v^2 - \vec{\pi}^2} \right ) && \, + \hspace{0.12em} F_T \hspace{0.62em} \left ( 1 + \frac{\pi_3^2}{v^2 - \vec{\pi}^2} \right )^2 \left ( 1 - \frac{\vec{\pi}^2}{v^2} \right ) \, .
\end{alignat}%
\endgroup
One can then compute the associated Levi-Civita connection and Riemann curvature tensor, whose detailed expressions are collected in \cref{app:geom}. Here, we summarize some salient properties:
\begin{itemize}
    \item When $F_{2D} \rightarrow 0$, we have $\tensor{\Gamma}{^h_{hh}} \rightarrow K' / 2K$ and $\tensor{\Gamma}{^{\pi_A}_{hh}} \rightarrow 0$, with primes indicating derivatives in $h$.

    \item More generally, we still have $\tensor{\Gamma}{^{\pi_B}_{hh}}, \, \tensor{\Gamma}{^{\pi_B}_{h \pi_3}}, \, \tensor{\Gamma}{^{\pi_B}_{\pi_3 \pi_3}} \rightarrow 0$ when $\pi_B \rightarrow 0$ for $B = 1, 2$, even if $F_{2D} \neq 0$. Additionally,
    \begin{equation}
        \tensor{\Gamma}{^{\pi_3}_{hh}} \rightarrow \frac{1}{2} \, \sqrt{1 - \frac{\pi_3^2}{v^2}} \, \frac{K' F_{2D} - 2 K F_{2D}'}{K(F_C + F_T) - F_{2D}^2} \, .
    \end{equation}

    \item There are four independent sectional curvatures. At the vacuum, the curvature components reduce to
    \begin{alignat}{3}
        & R_{h \pi_1 \pi_1 h} \, , \; && R_{h \pi_2 \pi_2 h} & \rightarrow &- \frac{K' F_C'}{4K} - \frac{F_C'^2}{4F_C} + \frac{F_C''}{2} \label{eq:scalar_curvature_1} \\
        & && &&- \left ( \frac{1}{v^2} + \frac{K' F_C'}{4 K^2} \right ) \frac{F_{2D}^2}{F_C} + \frac{F_C' \, F_{2D}^{\vphantom{'}} \, F_{2D}'}{2 K F_C} + \ldots \, , \notag \\
        & && R_{h \pi_3 \pi_3 h} & \rightarrow &- \frac{K' F_C'}{4K} - \frac{F_C'^2}{4 F_C} + \frac{F_C''}{2} \\
        & && &&- \left ( \frac{K'}{4K} + \frac{F_C'}{2 F_C} \right ) F_T' + \frac{F_T''}{2} - \frac{F_C'^2 \, F_T^2}{4 F_C^3} + \frac{F_C' \, F_T^{\vphantom{'}} \, F_T'}{2 F_C^2} - \frac{F_T'^2}{4 F_C} \notag \\
        & && &&- \left ( \frac{K'}{K} + \frac{F_C'}{F_C} \right ) \frac{F_C' F_{2D}^2}{4 K F_C} + \frac{F_C' \, F_{2D}^{\vphantom{'}} \, F_{2D}'}{2 K F_C} + \ldots \, , \notag \\
        & && R_{\pi_1 \pi_2 \pi_2 \pi_1} & \rightarrow & - \frac{F_C}{v^2} + \frac{F_C'^2}{4 K} + \frac{3 F_T}{v^2} + \frac{F_C'^2 \, F_{2D}^2}{4 K^2 F_C} + \ldots \, , \\
        & R_{\pi_1 \pi_3 \pi_3 \pi_1} \, , \enspace && R_{\pi_2 \pi_3 \pi_3 \pi_2} \; & \rightarrow & - \frac{F_C}{v^2} + \frac{F_C'^2}{4 K} - \frac{2 F_T}{v^2} + \frac{F_C' \, F_T'}{4K} - \frac{F_T^2}{v^2 F_C} + \frac{F_C'^2 \, F_{2D}^2}{4 K^2 F_C} + \ldots \, , \label{eq:scalar_curvature_4}
    \end{alignat}
    Omitted terms are at least cubic in $F_T$, $F_{2D}$ or their derivatives. The difference between $R_{h \pi_A \pi_A h}$ for $A = 3$ and for $A = 1, 2$ is due to the custodial symmetry violating terms $F_T$ and $F_{2D}$, and likewise for $R_{\pi_A \pi_B \pi_B \pi_A}$.
\end{itemize}

Consider the case $F_{2D} = 0$. The line $\mathcal{H}$ passing through the vacuum and parameterized by $\bm{h}$ at unit speed via
\begin{equation}
    \bm{h} = \int_0^h d\eta \; \sqrt{K(\eta)} \, , \quad \vec{\pi} = \vec{0} \, ,
\end{equation}
will then satisfy the geodesic equation
\begin{equation}
    \frac{d^2 \phi^I}{d \bm{h}^2} + \tensor{\Gamma}{^I_{JK}} \, \frac{d\phi^J}{d\bm{h}} \frac{d\phi^K}{d\bm{h}} = 0 \, .
\end{equation}
The orthonormal frame
\begin{equation}
    \bigg \{ \partial_{\bm{h}} \, , \, \partial_{\bm{\pi_1}} \, , \, \partial_{\bm{\pi_2}} \, , \, \partial_{\bm{\pi_3}} \bigg \} = \bigg \{ \frac{\partial_h}{\sqrt{K}} \, , \, \frac{\partial_{\pi_1}}{\sqrt{F_C}} \, , \, \frac{\partial_{\pi_2}}{\sqrt{F_C}} \, , \, \frac{\partial_{\pi_3}}{\sqrt{F_C + F_T}} \bigg \} \, ,
\end{equation}
is covariantly constant along $\mathcal{H}$, the first of which is its tangent vector. Sending out secondary geodesics in the $\partial_{\bm{\pi_A}}$ directions, we obtain a Fermi normal coordinate system $\phi^{\bm{I}}$ adapted to the primary geodesic $\mathcal{H}$. This corresponds to a field redefinition
\begin{align}
    h &= \frac{1}{K^{1/2}} \, \bm{h} - \frac{K'}{4 K^2} \, \bm{h}^2 + \frac{F_C'}{4 K F_C} \left ( \bm{\pi}_1^2 + \bm{\pi}_2^2 \right ) + \frac{F_C' + F_T'}{4 K (F_C + F_T)} \, \bm{\pi}_3^2 + \ldots \, , \\
    \pi_B &= \frac{1}{{F_C}^{1/2}} \, \bm{\pi_B} - \frac{F_C'}{2 K^{1/2} {F_C}^{3/2}} \, \bm{h} \bm{\pi_B} - \frac{\epsilon_{AB3} \, F_T}{v {F_C}^{3/2} (F_C + F_T)^{1/2}} \, \bm{\pi_A} \bm{\pi_3} + \ldots \, , \\
    \pi_3 &= \frac{1}{(F_C + F_T)^{1/2}} \, \bm{\pi_3} - \frac{F_C' + F_T'}{2 K^{1/2} (F_C + F_T)^{3/2}} \, \bm{h} \bm{\pi_3} + \ldots \, ,
\end{align}
about the vacuum, where $B = 1$ or $2$, $\epsilon$ is the Levi-Civita symbol and terms at first non-trivial order are shown.\footnote{The first deviation from Riemann normal coordinates is $\phi^I \sim \tensor{\Gamma}{^I_{\bm{J} \bm{K}, \, \bm{L}}} \, \phi^{\bm{J}} \phi^{\bm{K}} \phi^{\bm{L}}$ at the next order \cite{Klein:2007xj}.} Each $\phi^{\bm{I}}$ is a canonically normalized version of $\phi^I$.

The line $\mathcal{H}$ is significant --- it is the coordinate axis of the physical Higgs boson $\bm{h}$ unconstrained by symmetry. Generally, for each value of $\bm{h}$, the geometry along the Goldstone boson directions $\bm{\pi_A}$ must realize the $SU(2)_L \times U(1)_Y$ symmetry and is hence fixed by the four values of $K$, $F_C$, $F_T$ and $F_{2D}$.\footnote{The last is zero in the case considered here.} At the vacuum, these numbers are accessible via \emph{e.g.} the four sectional curvatures listed above. This is in contrast with the $\bm{h}$ direction, which contains not four numbers' but four functions' worth of information. By setting up Fermi normal coordinates aligned with $\mathcal{H}$, we are best adapted to reconstruct their functional dependence and hence any distant geometric features at finite $\bm{h}$ encoded within.

In the general scenario that $F_{2D} \neq 0$, one can still probe the physical Higgs boson direction by sending out a geodesic from the vacuum with initial velocity $\partial_{\bm{h}}$. While $\pi_1$ and $\pi_2$ will remain at zero, mixing of $h$ with $\pi_3$ means that the latter will begin to grow as $\sim F_{2D}$. The geodesic will not deviate significantly from the $h$ coordinate axis near the vacuum, in view of experimental constraints on $F_{2D}$. However, interesting geometric features typically reside at large distances $h \sim v$, and their detailed reconstruction will require knowledge of custodial symmetry violation at higher orders in the EFT expansion.\footnote{One can alternatively use a coordinate system derived by performing Fermi-Walker transport on an orthonormal frame along the non-geodesic $h$ coordinate axis. However, like Riemann normal coordinates, it will not be locally flat along the axis.}

\subsection{Fermion sector}

The invariant operators in the fermion sector include the Yukawa interactions
\begin{equation}
    \mathcal{L} \supset - \frac{v}{\sqrt{2}} \, y(h) \left [ \overbar{Q}_L \, U \, Q_R \right ] - \frac{v}{2 \sqrt{2}} \, y_T(h) \left [ \overbar{Q}_L \, U \sigma_3 \, Q_R \right ] + \text{h.c.} \, ,
\end{equation}
the Hermitian kinetic terms
\begin{subequations}
\begin{alignat}{3}
    \mathcal{L} &\supset &&\frac{i}{2} \, k_L(h) \, \big [ \overbar{Q}_L \overlrarrow{\slashed{\partial}} Q_L \big ] &&+ \frac{i}{4} \, k_{LT}(h) \left [ \overbar{Q}_L \, T \, (\slashed{\partial} Q_L) - (\slashed{\partial} \overbar{Q}_L) \, T \, Q_L \right ] \\
    &\quad + \, &&\frac{i}{2} \, k_R(h) \, \big [ \overbar{Q}_R \overlrarrow{\slashed{\partial}} Q_R \big ] &&+ \frac{i}{4} \, k_{RT}(h) \left [ \overbar{Q}_R \, \sigma_3 \, (\slashed{\partial} Q_R) - (\slashed{\partial} \overbar{Q}_R) \, \sigma_3 \, Q_R \right ] \, ,
\end{alignat}
\end{subequations}
and the anti-Hermitian kinetic terms
\begin{subequations}
\begin{alignat}{3}
    \mathcal{L} &\supset &&\frac{i}{v} \, \omega_L(h) \left [ \overbar{Q}_L \, \gamma^\mu \, Q_L \right ] \partial_\mu h &&+ \frac{i}{2v} \, \omega_{LT}(h) \left [ \overbar{Q}_L \, \gamma^\mu T \, Q_L \right ] \partial_\mu h \\
    &\quad + \, &&i \, \varpi_L(h) \left [ \overbar{Q}_L \, \gamma^\mu V_\mu \, Q_L \right ] &&+ \frac{i}{2} \, \varpi_{LT1}(h) \left [ \overbar{Q}_L \, (\slashed{\partial} T) \, Q_L \right ] \\
    &\quad + &&\frac{i}{2} \, \varpi_{LT2}(h) \left [ \overbar{Q}_L \, \gamma^\mu \, Q_L \right ] \Tr \big [ V_\mu T \big ] &&+ \frac{i}{2} \, \varpi_{LT3}(h) \left [ \overbar{Q}_L \, \gamma^\mu T \, Q_L \right ] \Tr \big [ V_\mu T \big ] \notag \\
    &\quad + &&\frac{i}{v} \, \omega_R(h) \left [ \overbar{Q}_R \, \gamma^\mu \, Q_R \right ] \partial_\mu h &&+ \frac{i}{2v} \, \omega_{RT}(h) \left [ \overbar{Q}_R \, \gamma^\mu \sigma_3 \, Q_R \right ] \partial_\mu h \\
    &\quad + &&i \, \varpi_R(h) \left [ \overbar{Q}_R \, \overbar{U} \gamma^\mu V_\mu U \, Q_R \right ] &&+ \frac{i}{2} \, \varpi_{RT1}(h) \left [ \overbar{Q}_R \, \overbar{U} (\slashed{\partial} T) U \, Q_R \right ] \\
    &\quad + &&\frac{i}{2} \, \varpi_{RT2}(h) \left [ \overbar{Q}_R \, \gamma^\mu \, Q_R \right ] \Tr \big [ V_\mu T \big ] &&+ \frac{i}{2} \, \varpi_{RT3}(h) \left [ \overbar{Q}_R \, \gamma^\mu \sigma_3 \, Q_R \right ] \Tr \big [ V_\mu T \big ] \, . \notag
\end{alignat}
\end{subequations}
Note that the kinetic terms involve only fermions in left-handed or right-handed doublets, while the Yukawa terms necessarily span fermions of both chiralities. Like in the scalar sector, we expect the $y$ and $k$ operators labeled with $T$ to be small compared to their custodially symmetric counterparts near the vacuum. While not our focus, four-fermion operators and their enumeration pose no difficulties.

Computing geometric quantities on the vector bundle $\mathcal{E}$, we obtain the fermion mass matrix
\begin{alignat}{3}
    &M_{\bar{t}_L t_R} &&= M_{\bar{t}_R t_L}^* &&= \, \frac{\sqrt{v^2 - \vec{\pi}^2} + i \pi_3}{\sqrt{2}} \left ( y + \frac{1}{2} \, y_T \right ) \, , \\
    &M_{\bar{t}_L b_R} &&= M_{\bar{b}_R t_L}^* &&= \, \frac{i \pi_1 + \pi_2}{\sqrt{2}} \left ( y - \frac{1}{2} \, y_T \right ) \, , \\
    &M_{\bar{b}_L t_R} &&= M_{\bar{t}_R b_L}^* &&= \, \frac{i \pi_1 - \pi_2}{\sqrt{2}} \left ( y + \frac{1}{2} \, y_T \right ) \, , \\
    &M_{\bar{b}_L b_R} &&= M_{\bar{b}_R b_L}^* &&= \, \frac{\sqrt{v^2 - \vec{\pi}^2} - i \pi_3}{\sqrt{2}} \left ( y - \frac{1}{2} \, y_T \right ) \, .
\end{alignat}
The Hermitian metric reads
\begin{align}
    k_{\bar{t}_L t_L} &= \, k_L + \left ( \frac{1}{2} - \frac{\pi_1^2 + \pi_2^2}{v^2} \right ) k_{LT} \, , \\
    k_{\bar{t}_L b_L} = k_{\bar{b}_L t_L}^* &= \, - \left [ i \left ( \pi_1 + \frac{\pi_2 \pi_3}{\sqrt{v^2 - \vec{\pi}^2}} \right ) + \left ( \pi_2 - \frac{\pi_1 \pi_3}{\sqrt{v^2 - \vec{\pi}^2}} \right ) \right ] \sqrt{1 - \frac{\vec{\pi}^2}{v^2}} \, \frac{k_{LT}}{v} \, , \\
    k_{\bar{b}_L b_L} &= \, k_L - \left ( \frac{1}{2} - \frac{\pi_1^2 + \pi_2^2}{v^2} \right ) k_{LT} \, , \\
    k_{\bar{t}_R t_R} &= \, k_R + \frac{1}{2} \, k_{RT} \, , \\
    k_{\bar{b}_R b_R} &= \, k_R - \frac{1}{2} \, k_{RT} \, .
\end{align}
We again defer the detailed expressions of the connection and curvature to \cref{app:geom}, but will mention some notable features of the latter:
\begin{itemize}
    \item If $(\bar{p}, r)$ are both top or bottom components of doublets of the same chirality, then
    \begin{equation}
        R_{\bar{p} r h \pi_3} \vbar \neq 0 \quad \text{and} \quad R_{\bar{p} r \pi_1 \pi_2} \vbar \neq 0 \, . \label{eq:curv_first_set}
    \end{equation}
    If one of $(\bar{p}, r)$ is a top component and the other is a bottom component, then
    \begin{alignat}{5}
        &R_{\bar{p} r h \pi_1} \vbar &&= i R_{\bar{p} r h \pi_2} \vbar &&= - \, R_{\bar{r} p h \pi_1}^* \vbar &&= - \, i R_{\bar{r} p h \pi_2}^* \vbar &&\neq 0 \, , \label{eq:curv_second_set_1} \\
        &R_{\bar{p} r \pi_1 \pi_3} \vbar &&= i R_{\bar{p} r \pi_2 \pi_3} \vbar &&= - \, R_{\bar{r} p \pi_1 \pi_3}^* \vbar &&= - \, i R_{\bar{p} r \pi_2 \pi_3}^* \vbar &&\neq 0 \, . \label{eq:curv_second_set_2}
    \end{alignat}
    Otherwise, the curvature vanishes at the vacuum.

     \item Custodial symmetry forbids the $\omega$'s, \emph{i.e.} the couplings to the physical Higgs boson, from entering the curvature. More generally, the first set \cref{eq:curv_first_set} is indeed solely determined by the $\varpi$'s, \emph{i.e.} the couplings to the Goldstone bosons, at the vacuum. However, the $\omega$'s do arise at the vacuum in the second set, \crefrange{eq:curv_second_set_1}{eq:curv_second_set_2}, and also beyond the vacuum in the first set for $\omega_L$ and $\omega_{LT}$.
    
    \item Custodial symmetry entails that $R_{\bar{t}_L t_L I J} = - \, R_{\bar{b}_L b_L I J}$ within the first set of curvatures. This relation is violated by non-vanishing $k_{LT}$ or $\varpi_{LT2}$. For example,
    \begin{equation}
        R_{\bar{t}_L t_L h \pi_3} \vbar \supset \frac{{k_L^+}'}{k_L^+} \, \frac{\varpi_L}{v} - \frac{\varpi_{LT2}'}{v} \enspace \text{is not the negative of} \enspace
        R_{\bar{b}_L b_L h \pi_3} \vbar \supset - \frac{{k_L^-}'}{k_L^-} \, \frac{\varpi_L}{v} - \frac{\varpi_{LT2}'}{v} \, ,
    \end{equation}
    where $k_L^{\pm} \coloneqq k_L^{\vphantom{\pm}} \pm k_{LT}^{\vphantom{\pm}} / 2$. The same applies to the other doublet $L \rightarrow R$.
\end{itemize}

Writing $\omega_L^{\pm} = \omega_L^{\vphantom{\pm}} \pm \omega_{LT}^{\vphantom{\pm}} / 2$ and likewise for $L \rightarrow R$,\footnote{These combinations for custodial symmetry violation should not be confused with the $\omega^{\pm}$ in \cite{Derda:2024jvo} that defines the vector bundle connection \cref{eq:vec_bundle_conn}.} the fermionic Fermi normal coordinates are given by the field redefinition
\begin{align}
    t_R &= \left [ 1 - \frac{\omega_R^+}{k_R^+} \, \frac{h}{v} + \frac{\varpi_R + \varpi_{RT2} + \varpi_{RT3}}{k_R^+} \, \frac{\pi_3}{v} \right ] \frac{\bm{t_R}}{(k_R^+)^{1/2}} \\
    &\quad - \frac{i \varpi_R - \varpi_{RT1}}{(k_R^+ \, k_R^-)^{1/2}} \, \frac{i \pi_1 + \pi_2}{v} \, \frac{\bm{b_R}}{(k_R^+)^{1/2}} + \ldots \, , \notag \\
    t_L &= (R \rightarrow L) + \frac{1}{2} \, \frac{k_{LT}}{k_L^+} \, \frac{i \pi_1 + \pi_2}{v} \, \frac{\bm{b_L}}{(k_L^-)^{1/2}} + \ldots \, , \\
    b_R &= \left [ 1 - \frac{\omega_R^-}{k_R^-} \, \frac{h}{v} - \frac{\varpi_R - \varpi_{RT2} + \varpi_{RT3}}{k_R^-} \, \frac{\pi_3}{v} \right ] \frac{\bm{b_R}}{(k_R^-)^{1/2}} \\
    &\quad - \frac{i \varpi_R + \varpi_{RT1}}{(k_R^+ \, k_R^-)^{1/2}} \, \frac{i \pi_1 - \pi_2}{v} \, \frac{\bm{t_R}}{(k_R^-)^{1/2}} + \ldots \, , \notag \\
    b_L &= (R \rightarrow L) - \frac{1}{2} \, \frac{k_{LT}}{k_L^-} \, \frac{i \pi_1 - \pi_2}{v} \, \frac{\bm{t_L}}{(k_L^+)^{1/2}} + \ldots \, ,
\end{align}
at first non-trivial order in scalar coordinates about the vacuum. To be clear, the vector bundle connection coefficients do not affect the geodesic $\mathcal{H}$ on $\mathcal{M}$ used for Fermi normal coordinates.\footnote{If viewed from the perspective of the total space of the vector bundle, $\mathcal{H}$ lies on the null section $\chi^p = 0$ and the geodesic equation is trivially satisfied in the fermionic directions.}

\subsection{Imprints of new physics on field space}

The geometric reformulation of EFT operators in the SM Higgs sector provides a meaningful interpretation of searches for new physics, without ambiguity from field redefinitions --- constraints on Wilson coefficients in the functions $K, F_C, \ldots$ are truly bounds on field space tensors like $R$. Aside from the quantitative values of these tensors, there are more general qualitative features with deeper physical implications. We now consider how such geometric information can be recovered from scattering experiments, starting with those readily accessible at the vacuum and then venturing beyond.

To identify physics beyond the SM, one must establish the geometry of the SM itself. As has been well understood, the field space of the SM is flat \cite{Alonso:2015fsp}. This is evident once we substitute the only non-vanishing kinetic terms
\begin{equation}
    \text{SM:} \quad K = 1 \, , \quad F_C = \frac{r^2}{v^2} \, , \quad k_L = k_R = 1 \, ,
\end{equation}
with $r = v + h$ into the curvature expressions derived above. Hence, any new physics at the one- or two-derivative level is signaled by non-zero curvature.\footnote{Derivative field redefinitions may complicate the precise physical content encapsulated in the curvature.} At the zero-derivative level, the SM itself is already non-trivial. The scalar potential assumes a specific form 
\begin{equation}
    \text{SM:} \quad V = - \frac{\mu^2}{2} \, r^2 + \frac{\lambda}{4} \, r^4 \, ,
\end{equation}
and the fermion mass matrix is not invariant under $t \leftrightarrow b$ --- the $y$ and $y_T$ that enter $M$ are non-vanishing constants with $0 < | \, y_T / y \, | \ll 1$, representing small but non-zero custodial symmetry violation.

Any first signs of new physics will reside in four-point scattering amplitudes, which grant immediate access to the geometry at the vacuum. The relevant scalar amplitudes read
\begin{align}
    \mathcal{A} \, \Big ( \bm{\pi_A} \, \bm{\pi_A} \rightarrow \bm{h} \, \bm{h} \Big ) \supset &- R_{\bm{h} \bm{\pi_A} \bm{\pi_A} \bm{h}} \vbar \left ( E^2 - 2 \, m_{\bm{h}}^2 \right ) - V_{; \, \bm{\pi_A} \bm{\pi_A} \bm{h} \bm{h}} \vbar \, , \\
    \mathcal{A} \, \Big ( \bm{\pi_A} \, \bm{\pi_A} \rightarrow \bm{\pi_B} \, \bm{\pi_B} \Big ) \supset &- R_{\bm{\pi_A} \bm{\pi_B} \bm{\pi_B} \bm{\pi_A}} \vbar \left ( E^2 - \frac{2}{3} \, m_{\bm{\pi_A}}^2 - \frac{2}{3} \, m_{\bm{\pi_B}}^2 \right ) - V_{; \, (\bm{\pi_A} \bm{\pi_A} \bm{\pi_B} \bm{\pi_B})} \vbar \, .
\end{align}
Focusing on the large momentum limit, we display terms that do not diminish with energy, and consider the Goldstone bosons $\bm{\pi_{1/2}}$ and $\bm{\pi_3}$ equivalent to the longitudinal modes of the $W^{\pm}$ and $Z^0$ bosons. Converting between Fermi and CCWZ curvature components at the vacuum amounts to appending appropriate wavefunction renormalization factors, due to the tensorial transformation laws
\begin{equation}
    R_{\bm{h} \bm{\pi_3} \bm{\pi_3} \bm{h}} \vbar = \left ( \frac{\partial \phi^I}{\partial \bm{h}} \, \frac{\partial \phi^J}{\partial \bm{\pi_3}} \, \frac{\partial \phi^K}{\partial \bm{\pi_3}} \, \frac{\partial \phi^L}{\partial \bm{h}} \, R_{IJKL} \right ) \biggvbar = \frac{R_{h \pi_3 \pi_3 h}}{K (F_C + F_T)} \biggvbar \, ,
\end{equation}
and so on. Evidently, any growth of the scalar amplitudes with energy implies that the scalar field space $\mathcal{M}$ is curved at the vacuum and deviates from the SM. Furthermore, if there is a difference in the rate of growth for amplitudes involving $\bm{\pi_{1/2}}$ and $\bm{\pi_3}$, there must then be a new source of custodial symmetry violation beyond the SM, rendering $\mathcal{M}$ anisotropic \cite{Alonso:2016oah}. According to \crefrange{eq:scalar_curvature_1}{eq:scalar_curvature_4}, this difference is first- and second-order in $F_T$ and $F_{2D}$ respectively, and hence a leading indicator of a CP-even origin.

Writing $P_L = (1 - \gamma_5) / 2$ for the left-handed projection operator, the relevant fermion amplitudes include
\begin{align}
    \mathcal{A} \, \Big ( \bm{\bar{t}_L} \, \bm{t_L} \rightarrow \bm{\pi_3} \, \bm{h} \Big ) \supset &- \overbar{v}(q_1)  \left [ R_{\bm{\bar{t}_L} \bm{t_L} \bm{h} \bm{\pi_3}} \vbar \, \slashed{q}_4 + M_{\bm{\bar{t}_L} \bm{t_L}; \, \bm{\pi_3} \bm{h}} \vbar \right ] P_L \, u(q_2) \, , \\
    \mathcal{A} \, \Big ( \bm{\bar{t}_L} \, \bm{t_L} \rightarrow \bm{\pi_1} \, \bm{\pi_2} \Big ) \supset &- \overbar{v}(q_1) \left [ \frac{1}{2} \, R_{\bm{\bar{t}_L} \bm{t_L} \bm{\pi_1} \bm{\pi_2}} \vbar \left ( \slashed{q}_3 - \slashed{q}_4 \right ) + M_{\bm{\bar{t}_L} \bm{t_L}; \, (\bm{\pi_1}  \bm{\pi_2})} \vbar \right ] P_L \, u(q_2) \, ,
\end{align}
and $t_L \rightarrow b_L$, which probe the first set of curvatures on the vector bundle $\mathcal{E}$. Like before, any quadratic growth of the amplitudes with energy is an indicator of non-zero curvature and hence new physics. Moreover, a difference in the cross sections under $t_L \leftrightarrow b_L$ may signal a new source of custodial symmetry violation, rendering $\mathcal{E}$ asymmetric in the top and bottom directions. In the high energy limit, the violation can be unequivocally attributed to non-zero $k_{LT}$ or $\varpi_{LT2}$, while violation at lower energies must be in excess of the SM Yukawa couplings. Between the two choices of final states, scattering into $\bm{\pi_1} \bm{\pi_2}$ also gives access to the value of $\varpi_{LT1}$ at the vacuum. There are additionally the amplitudes
\begin{equation}
    \mathcal{A} \, \Big ( \bm{\bar{t}_L} \, \bm{b_L} \rightarrow \bm{\pi_B} \, \bm{h} \Big ) \supset - \overbar{v}(q_1)  \left [ R_{\bm{\bar{t}_L} \bm{b_L} \bm{h} \bm{\pi_B}} \vbar \, \slashed{q}_4 + M_{\bm{\bar{t}_L} \bm{b_L}; \, \bm{\pi_B} \bm{h}} \vbar \right ] P_L \, u(q_2) \, ,
\end{equation}
for $B = 1, 2$ and $t_L \leftrightarrow b_L$, which at high energies assess the second set of curvatures and the derivative Higgs coupling $\omega_{LT}$ that appear within. The same analysis holds for $L \rightarrow R$.

While amplitudes at low points provide an immediate assessment of the geometry at the vacuum and diagnose any anisotropy or asymmetry due to custodial symmetry violation, we expect that amplitudes at higher points begin to reveal any curvature singularity beyond the vacuum left behind by heavy states. This has been discussed in detail in the previous section for one of the vector bundle curvatures that enter the fermion amplitude \cref{eq:A_ttpi} as the sole contact term; if it dominates, then the high-energy behavior of the amplitude is in direct correspondence with the Taylor coefficients of the curvature at the vacuum in Fermi normal coordinates, allowing us to estimate its behavior a finite distance away. Generally speaking, extracting other field space curvatures from their corresponding amplitudes is more difficult because of additional competing contact terms on top of factorizable ones. For the Higgs sector specifically, it turns out that $F_{2D} = 0$ entails
\begin{equation}
    R_{\bm{h} \bm{\pi_A} \bm{\pi_B} \bm{h}} \vbar_{\mathcal{H}} \propto \delta_{\bm{A} \bm{B}} \, , \enspace R_{\bm{\pi_A} \bm{\pi_B} \bm{\pi_C} \bm{h}} \vbar_{\mathcal{H}} \propto (3 \delta_{\bm{C} \bm{3}} - 1) \, \epsilon_{\bm{A} \bm{B} \bm{C}} \enspace \text{and} \enspace R_{\bm{h} (\bm{\pi_1} \bm{\pi_2}| \bm{h}; \, |\bm{\pi_3})} \vbar_{\mathcal{H}} = 0 \, ,
\end{equation}
on the Higgs coordinate axis $\mathcal{H}$ which is a geodesic. These identities can help eliminate some competing contact terms for certain theories. Consider the three-Goldstone amplitude
\begin{align}
    &\mathcal{A} \, \Big ( \bm{\pi_A} \, \bm{\pi_B} \rightarrow \bm{\pi_C} \, \bm{h}^{N+1} \Big ) \supset - \, \tensor{R}{_{\bm{h} (\bm{\pi_A} \bm{\pi_B}| \bm{h}; \, |\bm{\pi_C}) \bm{h}^{N-1}}} \vbar \, s_{123} \\
    &\hspace{5em} - \frac{2}{3} \, \tensor{R}{_{\bm{\pi_C} (\bm{\pi_A} \bm{\pi_B}) \bm{h}; \, \bm{h}^N}} \vbar \, s_{12} - \frac{2}{3} \, \tensor{R}{_{\bm{\pi_B} (\bm{\pi_A} \bm{\pi_C}) \bm{h}; \, \bm{h}^N}} \vbar \, s_{13} - \frac{2}{3} \, \tensor{R}{_{\bm{\pi_A} (\bm{\pi_B} \bm{\pi_C}) \bm{h}; \, \bm{h}^N}} \vbar \, s_{23} \notag \\
    &\hspace{5em} - \left ( \tensor{R}{_{\bm{\pi_A} \bm{\pi_B} \bm{\pi_C} \bm{h} \vphantom{\bm{h}^{N-2}}}} \vbar \, s_{14} + \tensor{R}{_{\bm{\pi_C} \bm{\pi_B} \bm{\pi_A} \bm{h} \vphantom{\bm{h}^{N-2}}}} \vbar \, s_{34} \right ) \left ( \tensor{R}{_{\bm{h} \bm{\pi_B} \bm{\pi_B} \bm{h}; \, \bm{h}^{N-2}}} \vbar \, \frac{s_{1234}}{s_{134}} \right ) + \ldots \, , \notag
\end{align}
written in Fermi normal coordinates along $\mathcal{H}$. The contact term in the first line vanishes if $F_{2D} = 0$, while competing factorizable terms in the third line all depend on $R_{\bm{h} \bm{\pi_D} \bm{\pi_D} \bm{h}}$ on $\mathcal{H}$; for example, the diagram explicitly shown has a propagator of the same flavor as the lone Goldstone external leg $\bm{\pi_B}$ at one of its two ends. Should $R_{\bm{h} \bm{\pi_D} \bm{\pi_D} \bm{h}}$ be small relative to $R_{\bm{\pi_D} \bm{\pi_E} \bm{\pi_E} \bm{\pi_D}}$, the simplified amplitude
\begin{equation}
    \mathcal{A} \, \Big ( \bm{\pi_1} \, \bm{\pi_2} \rightarrow \bm{\pi_3} \, \bm{h}^{N+1} \Big ) \approx \frac{1}{2} \, \tensor{R}{_{\bm{\pi_1} \bm{\pi_2} \bm{\pi_3} \bm{h}; \, \bm{h}^N}} \vbar \left ( s_{13} - s_{23} \right ) \, , \label{eq:A_pi3_simple}
\end{equation}
becomes a direct probe of its corresponding curvature component at high energies.\footnote{These curvature components are not all independent, and can be converted into one another via a polarization identity. It is not surprising that neglecting some curvature components makes the amplitude depend on fewer curvature components. The simplicity of the dependence in \cref{eq:A_pi3_simple} is less trivial.} We can then also expect a simple relationship \cref{eq:singularity_from_unitarity} between the scale of unitarity violation at moderate multiplicity $N$ and any singularity of $R_{\bm{\pi_1} \bm{\pi_2} \bm{\pi_3} \bm{h}}$ along $\mathcal{H}$. Roughly speaking, theories for which such factorizable terms drop out must be more strongly curved in the $S^3$ along the Goldstone directions than in the $\mathbb{R}$ along the Higgs direction. Outside of these situations, numerical techniques may be required to precisely separate all competing contributions, contact and factorizable \cite{Cohen:2021ucp}. Nevertheless, by rendering $\mathcal{H}$ a geodesic, $F_{2D} = 0$ remains a useful condition that aligns scattering processes with an increasing number of Higgs states directly with the geometry along the Higgs axis. Otherwise, when measuring any proper distance with respect to the vacuum, one must account for deviations that are hard to nail down with perturbative amplitudes.

For concreteness, consider extending the SM with a real $SU(2)_L$ triplet scalar $\Phi_A$:  
\begin{align}
    \mathcal{L}_{\text{UV}} &\supset - \, V_{\text{UV}} + | \partial H |^2 + \frac{1}{2} \, ( \partial \Phi )^2 \, , \\
    V_{\text{UV}} &= - \, \mu^2 \, |H|^2 + \lambda \, |H|^4 - \frac{\mu_{\Phi}^2}{2} \, \Phi^2 + \frac{\lambda_{\Phi}}{4} \, \Phi^4 - \kappa \, |H|^2 \, \Phi^2 - \eta \, \overbar{H} \sigma_A H \, \Phi_A \, ,
\end{align}
where we parameterize $\Phi_A$ in the unitary basis as \cite{Cohen:2020xca}
\begin{equation}
    \vec{\,\Phi} = \frac{2f}{r^2} \, \exp \begin{pmatrix}
        0 & 0 & \beta_2 \\ 0 & 0 & - \beta_1 \\ - \beta_2 & \beta_1 & 0
    \end{pmatrix} \overbar{H} \vec{\sigma} H \, .
\end{equation}
This field leaves the fermion sector untouched, but introduces new custodial symmetry violation in the scalar sector due to the coupling $\eta$. We take the constants to satisfy $\lambda, \lambda_{\Phi}, \eta > 0$ and $|\kappa| < \sqrt{\lambda \lambda_\Phi}$, ensuring in particular that $V_{\text{UV}}$ is bounded below and that the cubic polynomial
\begin{equation}
    - \frac{\eta}{2} \, r^2 - (\kappa r^2 + \mu_{\Phi}^2) \, f + \lambda_{\Phi} \, f^3 = 0 \, ,
\end{equation}
has a positive solution $f(r) > 0$. Following \cite{Cohen:2020xca}, we integrate out $\Phi_A$ at tree level to get the EFT\footnote{Compared to Eq.~(7.47) in \cite{Cohen:2020xca}, we find the opposite sign for the coefficient of the $(\partial |H|^2)^2$ term; as a result $K$ depends only on $f'$ and not $f$.}
\begin{align}
    \mathcal{L} \supset &- V + \frac{1}{2} \, (1 + f'^2) \, \partial_\mu h \, \partial^\mu h + \frac{1}{4} \, (r^2 + 4f^2) \Tr \left [ V_\mu V^\mu \right ] - \frac{1}{2} \, f^2 \Tr \left [ V_\mu T \right ] \Tr \left [ V^\mu T \right ] \, , \\
    V = &- \frac{\mu^2}{2} \, r^2 + \frac{\lambda}{4} \, r^4 - \frac{\mu_{\Phi}^2}{2} \, f^2 + \frac{\lambda_{\Phi}}{4} \, f^4 - \frac{\kappa}{2} \, r^2 f^2 - \frac{\eta}{2} \, r^2 f \, .
\end{align}
Its kinetic terms read
\begin{equation}
    K = 1 + f'^2 \, , \quad F_C = \frac{r^2 + 4f^2}{v^2} \, , \quad F_T = - \frac{4f^2}{v^2} \, , \quad F_{2D} = 0 \, , \quad k_L = k_R = 1 \, ,
\end{equation}
and determine the field space curvatures to be
\begin{alignat}{3}
    &R_{\bm{h} \bm{\pi_1} \bm{\pi_1} \bm{h}} \vbar_{\mathcal{H}} & = \; & R_{\bm{h} \bm{\pi_2} \bm{\pi_2} \bm{h}} \vbar_{\mathcal{H}} &&= \frac{1}{(1 + f'^2) \, (r^2 + 4f^2)} \left [ \frac{(4f - rf') \, f''}{1 + f'^2} + \frac{4(f - rf')^2}{r^2 + 4f^2} \right ] \, , \\
    & && R_{\bm{h} \bm{\pi_3} \bm{\pi_3} \bm{h}} \vbar_{\mathcal{H}} &&= - \, \frac{f' f''}{r \, (1 + f'^2)^2} \, , \\
    & && R_{\bm{\pi_1} \bm{\pi_2} \bm{\pi_2} \bm{\pi_1}} \vbar_{\mathcal{H}} &&= - \, \frac{(4f - rf')^2}{(1 + f'^2) \, (r^2 + 4f^2)^2} \, , \\
    & R_{\bm{\pi_1} \bm{\pi_3} \bm{\pi_3} \bm{\pi_1}} \vbar_{\mathcal{H}} & \; = \; & R_{\bm{\pi_2} \bm{\pi_3} \bm{\pi_3} \bm{\pi_2}} \vbar_{\mathcal{H}} &&= \frac{1}{r^2 + 4f^2} \left [ \frac{r + 4ff'}{r \, (1 + f'^2)} - \frac{r^2}{r^2 + 4f^2} \right ] \, , \\
    & && R_{\bm{\bar{p}} \bm{r} \bm{I} \bm{J}} \vbar_{\mathcal{H}} &&= \, 0 \, ,
\end{alignat}
where we display their expressions restricted to $\mathcal{H}$ for brevity. In particular, at the vacuum $\bm{h} = 0$ on $\mathcal{H}$, we see that the four-point scalar amplitudes will identify a non-zero $F_T$ from the differences between $\bm{\pi_{1/2}}$ and $\bm{\pi_3}$, while the fermionic ones will remain unchanged from the SM. This corroborates with the source of new physics, a CP-even heavy field that couples not to the SM fermions, but to the SM scalars and with additional custodial symmetry violation. Away from $\bm{h} = 0$, the triplet generates singularities on $\mathcal{M}$ of the form \cref{eq:order-d-pole} that arise when $K$, $F_C$, $F_C + F_T$ or the discriminant of the cubic polynomial vanishes. We can hence expect scalar amplitudes with a larger but finite number of Higgs states to be sensitive to these singularities as measured along the geodesic $\mathcal{H}$. This is subject to the various complications already discussed, with some possible simplifications when $\eta$ and $\mu_\Phi$ are small and $f$ becomes mostly linear in $r$, so that the four-Goldstone curvatures dominate the two-Goldstone ones. Meanwhile, higher-point fermionic amplitudes will not grow with energy just like in the SM, respecting perturbative unitarity and reflecting the absence of singularities in the curvatures on $\mathcal{E}$.

\section{Conclusion and outlook}
\label{sec:conclusion}

Field space geometry is an important theoretical framework for parameterizing the space of effective field theories, and a useful practical tool for extracting physical meaning from experimental measurements. This work refines the general geometric construction for scalar and fermionic degrees of freedom and develops further machinery that clarifies the geometric implications of physical observables. Concretely, we supply a fully covariant formulation of fermionic field space as a vector bundle, and establish Fermi normal coordinates as a direct link between scattering amplitudes and geometric features along a geodesic. We then focus on the Higgs sector and impose the necessary symmetries on the geometry. Attention is given to custodial symmetry violation, the geometric deformations it introduces in the Goldstone boson and fermion directions, and the complications it may bring to geometric reconstruction in the Higgs boson direction. Overall, this work serves to bridge existing gaps in geometric formalism and align it closer with phenomenological applications.

An avenue for future work is the incorporation of more particles. Leptons or multiple fermion generations can be readily accommodated in the present framework, and any interesting flavor structure will arise as non-diagonality in the fermionic directions. Meanwhile, the general treatment of gauge bosons requires extensions of field space with additional degrees of freedom. Other extensions in formalism include the geometric treatment of more general field redefinitions \cite{Cohen:2022uuw, Cheung:2022vnd, Craig:2023wni, Craig:2023hhp, Alminawi:2023qtf, Cohen:2023ekv, Cohen:2024bml, Lee:2024xqa, Cohen:2024fak}, loop effects \cite{Alonso:2022ffe, Helset:2022pde, Assi:2023zid, Jenkins:2023rtg, Jenkins:2023bls, Gattus:2024ird, Li:2024ciy, Aigner:2025xyt, Cohen:2025dex, Assi:2025fsm} or other EFT operators \cite{Assi:2023zid, Derda:2024jvo, Lee:2024xqa, Assi:2025fsm}, which have been explored in the literature to various extents. Access to geometry is no less important than the geometry itself, and experimentally viable probes of field space also deserve further study, such as scattering amplitudes and their intertwining of field space with kinematics \cite{Cheung:2021yog, Derda:2024jvo, Cohen:2025dex}. The possibility of more exotic geometric features away from the perturbative vacuum and their imprints on physical observables \cite{Alonso:2023jsi, Cheung:2024wme} suggest alternative prospects for physics beyond the Standard Model and opportunities for discovery.

\newpage
\acknowledgments

The authors are grateful for useful conversations with Andrew Fee, Julie Pagès, Chia-Hsien Shen and Dave Sutherland, and valuable comments from Timothy Cohen, Julie Pagès, Chia-Hsien Shen, Dave Sutherland and Zhengkang Zhang. IKL thanks the University of California Education Abroad Program for supporting his visit to the University of California, Santa Barbara. YTL thanks National Taiwan University for hospitality during the completion of this work. This work was supported in part by the U.S. Department of Energy under the grant DE-SC0011702 and performed in part at the Kavli Institute for Theoretical Physics, supported by the National Science Foundation under the grant NSF PHY-1748958.

\appendix
\section{A formula for phase space integration}
\label{app:phase_space}

Given $N+2$ distinguishable massless states, we derive the closed-form formula \cref{eq:phase_space_integral}, with $\rho_{N+2}$ and $\alpha_{N+2}$ set to unity, for the phase space integral of the product of two of their momenta. The form
\begin{equation}
    \int d\mathrm{LIPS} \; q_{\bm{I_J}}^\mu \, q_{\bm{I_K}}^{\vphantom{\mu} \nu} = E^{2N} \left ( a_{N+2} \, Q^\mu \, Q^\nu + b_{N+2} \, E^2 \,\eta^{\mu\nu} \right ) \, ,
\end{equation}
is fixed by Lorentz symmetry and dimensional analysis, with constants $a_{N+2}$ and $b_{N+2}$ to be determined. We take $\bm{I_J}$ and $\bm{I_K}$ to be among the first two states without loss of generality. Using the identity \cite{Kleiss:1985gy}
\begin{equation}
    1 = \int d^4 L \; \delta^{(4)} \left ( L^\rho - \sum_{\bm{J} = \bm{1}}^{\bm{N+1}} q_{\bm{I_J}}^\rho \right ) \int dW^2 \; \delta \left (L^2 - W^2 \right ) \, ,
\end{equation}
we decompose the $(N+2)$-state integral as
\begin{multline}
    E^{2N} \left ( a_{N+2} \, Q^\mu \, Q^\nu + b_{N+2} \, E^2 \,\eta^{\mu\nu} \right ) \\ = \int_0^{E^2} \frac{dW^2}{2 \pi} \int d\mathrm{LIPS}_2 \; W^{2(N-1)} \left ( a_{N+1} \, L^\mu \, L^\nu + b_{N+1} \, W^2 \,\eta^{\mu\nu} \right ) \, ,
\end{multline}
where
\begin{equation}
    d\mathrm{LIPS}_2 = \frac{d^3 q_{\bm{I_{N+2}}}}{(2\pi)^3 \, 2 E_{\bm{I_{N+2}}}} \times d^4 L \times 2 \pi \times \delta \left ( L^2 - W^2 \right ) \times \delta^{(4)} \left ( Q^\rho - L^\rho - q_{\bm{I_{N+2}}}^\rho \right ) \, .
\end{equation}
The variables $L$ and $W$ represent the total momentum and energy of the first $N+1$ states. We now have a two-state integral, over $L$ and $q_{\bm{I_{N+2}}}$, of an $(N+1)$-state integral of the same form. Evaluating the two-state integral yields the recursion relations
\begin{align}
    a_{N+2} &= \frac{a_{N+1}}{16 \pi^2 N (N+3)} \, , \\
    b_{N+2} &= \frac{b_{N+1}}{16 \pi^2 (N+1) (N+2)} - \frac{a_{N+1}}{32 \pi^2 N (N+1) (N+2) (N+3)} \, ,
\end{align}
with the solutions
\begin{equation}
    a_{N+2} = \frac{6 \, a_0}{(4 \pi)^{2N} N! \, (N+3)!} \, , \quad b_{N+2} = \frac{(2N + 6) \, b_0 - N a_0}{(4 \pi)^{2N} (N+1)! \, (N+3)!} \, .
\end{equation}
The seeds
\begin{equation}
    \left (a_0, b_0 \right ) = \begin{cases}
        \left ( + \frac{1}{24 \pi}, \, - \frac{1}{96 \pi} \right )  &\text{if } \bm{J} = \bm{K} \\
        \left ( + \frac{1}{48 \pi}, \, + \frac{1}{96 \pi} \right ) &\text{otherwise}
    \end{cases} \, ,
\end{equation}
at $N = 0$ then give the desired result.

\section{A compilation of geometric expressions}
\label{app:geom}

Calculating the field space geometry of the SM Higgs sector is straightforward in principle but rather unwieldy in practice. We begin with the scalar base space $\mathcal{M}$. For our purposes, it suffices to take the Goldstone boson coordinates $\pi_A$ and the custodial symmetry violating terms $F_T$ and $F_{2D}$ to be small --- after all, scattering amplitudes grant us access primarily to the vacuum. With this simplification and the shorthands
\begin{equation}
    \xi_A, \, \zeta_A = \begin{cases}
        \phantom{-} \, \pi_2 / v &\text{for } A = 1 \\
        - \, \pi_1 / v &\text{for } A = 2 \\
        - \, 1, \, 0 &\text{for } \xi_3, \, \zeta_3
    \end{cases}
    \quad \text{and} \quad
    c_{AB} = \begin{cases}
        \phantom{-} \, 2 &\text{if } (A,B) = (1,3) \text{ or } (2,3) \\
        - \, 2 &\text{if } (A,B) = (1,2) \text{ or } (2,1) \\
        \phantom{-} \, 0 &\text{otherwise}
    \end{cases} \, ,
\end{equation}
we list the leading terms in each Levi-Civita connection coefficient:
\begin{alignat}{6}
    &\tensor{\Gamma}{^h_{hh}} &&\supset \frac{K'}{2K} \, , && \tensor{\Gamma}{^h_{h \pi_A}} &&\supset - \frac{F_C' \, F_{2D}^{\vphantom{'}}}{2 K F_C} \, \xi_A \, , \qquad && \tensor{\Gamma}{^{\pi_A}_{hh}} &&\supset - \frac{K' F_{2D}^{\vphantom{'}} - 2 K^{\vphantom{'}} F_{2D}'}{2 K F_C} \, \xi_A \, , \\
    &\tensor{\Gamma}{^h_{\pi_1 \pi_1}} &&\supset - \frac{F_C'}{2K} \, , && \tensor{\Gamma}{^h_{\pi_2 \pi_2}} &&\supset - \frac{F_C'}{2K} \, , &&\tensor{\Gamma}{^h_{\pi_3 \pi_3}} &&\supset - \frac{F_C' + F_T'}{2K} \, , \\
    &\tensor{\Gamma}{^h_{\pi_1 \pi_2}} &&\supset \frac{F_T'}{2K} \, \zeta_3 \, , \qquad && \tensor{\Gamma}{^h_{\pi_2 \pi_3}} &&\supset \frac{F_T'}{2K} \, \zeta_2 \, , && \tensor{\Gamma}{^h_{\pi_3 \pi_1}} &&\supset \frac{F_T'}{2K} \, \zeta_1 \, , \\
    &\tensor{\Gamma}{^{\pi_1}_{h \pi_1}} &&\supset \frac{F_C'}{2 F_C} \, , && \tensor{\Gamma}{^{\pi_2}_{h \pi_2}} &&\supset \frac{F_C'}{2 F_C} \, , && \tensor{\Gamma}{^{\pi_3}_{h \pi_3}} &&\supset \frac{F_C' + F_T'}{2 (F_C + F_T)} \, ,
\end{alignat}
\begin{align}
    \tensor{\Gamma}{^{\pi_1}_{h \pi_2}} &\supset \frac{F_C' \, F_T^{\vphantom{'}} - F_C^{\vphantom{'}} \, F_T'}{2 F_C^2} \, \zeta_3 - \frac{F_{2D}}{v F_C} \, \xi_3 \, , \quad & \tensor{\Gamma}{^{\pi_2}_{h \pi_1}} &\supset \frac{F_C' \, F_T^{\vphantom{'}} - F_C^{\vphantom{'}} \, F_T'}{2 F_C^2} \, \zeta_3 + \frac{F_{2D}}{v F_C} \, \xi_3 \, , \\
    \tensor{\Gamma}{^{\pi_2}_{h \pi_3}} &\supset \frac{F_C' \, F_T^{\vphantom{'}} - F_C^{\vphantom{'}} \, F_T'}{2 F_C^2} \, \zeta_2 - \frac{F_{2D}}{v F_C} \, \xi_1 \, , & \tensor{\Gamma}{^{\pi_3}_{h \pi_2}} &\supset \frac{F_C' \, F_T^{\vphantom{'}} - F_C^{\vphantom{'}} \, F_T'}{2 F_C^2} \, \zeta_2 + \frac{F_{2D}}{v F_C} \, \xi_1 \, , \\
    \tensor{\Gamma}{^{\pi_3}_{h \pi_1}} &\supset \frac{F_C' \, F_T^{\vphantom{'}} - F_C^{\vphantom{'}} \, F_T'}{2 F_C^2} \, \zeta_1 - \frac{F_{2D}}{v F_C} \, \xi_2 \, , & \tensor{\Gamma}{^{\pi_1}_{h \pi_3}} &\supset \frac{F_C' \, F_T^{\vphantom{'}} - F_C^{\vphantom{'}} \, F_T'}{2 F_C^2} \, \zeta_1 + \frac{F_{2D}}{v F_C} \, \xi_2 \, ,
\end{align}
\begin{flalign}
    \hspace{0.85em} \tensor{\Gamma}{^{\pi_A}_{\pi_B \pi_B}} \supset \frac{F_C' \, F_{2D}^{\vphantom{'}}}{2 K F_C} \, \xi_A + \frac{F_C + c_{AB} \, F_T}{v F_C} \, \frac{\pi_A}{v} \, , &&
\end{flalign}
\begingroup
\allowdisplaybreaks
\begin{align}
    \tensor{\Gamma}{^{\pi_1}_{\pi_1 \pi_2}} &\supset \frac{F_T}{v F_C} \, \zeta_1 \, , \qquad & \tensor{\Gamma}{^{\pi_2}_{\pi_2 \pi_3}} &\supset \frac{F_T}{v F_C} \, \zeta_3 \, , \qquad & \tensor{\Gamma}{^{\pi_3}_{\pi_3 \pi_1}} &\supset \frac{F_T}{v F_C} \, \zeta_2 \, , \\
    \tensor{\Gamma}{^{\pi_1}_{\pi_1 \pi_3}} &\supset - \frac{F_T}{v F_C} \, \zeta_3 \, , & \tensor{\Gamma}{^{\pi_2}_{\pi_2 \pi_1}} &\supset - \frac{F_T}{v F_C} \, \zeta_2 \, , & \tensor{\Gamma}{^{\pi_3}_{\pi_3 \pi_2}} &\supset - \frac{F_T}{v F_C} \, \zeta_1 \, , \\
    \tensor{\Gamma}{^{\pi_1}_{\pi_2 \pi_3}} &\supset \frac{F_T}{v F_C} \, \xi_3 \, , & \tensor{\Gamma}{^{\pi_2}_{\pi_3 \pi_1}} &\supset - \frac{F_T}{v F_C} \, \xi_3 \, , & \tensor{\Gamma}{^{\pi_3}_{\pi_1 \pi_2}} &\supset \frac{F_T}{v F_C} \, \zeta_3 \, ,
\end{align}%
\endgroup
where an entry of $\zeta_3$ means that the leading term is quadratic in $\pi_A$. Working to this order yields the curvature tensor strictly at the vacuum, which has been listed in \crefrange{eq:scalar_curvature_1}{eq:scalar_curvature_4}. We note that should $F_{2D}$ happen to vanish, the curvature tensor along the entire coordinates axis $\mathcal{H}$ of $h$ has the compact expression
\begin{alignat}{3}
    &R_{h \pi_1 \pi_1 h} \vbar_{\mathcal{H}} & = \; & R_{h \pi_2 \pi_2 h} \vbar_{\mathcal{H}} &&= - \frac{K' F_C'}{4 K} - \frac{F_C'^2}{4 F_C} + \frac{F_C''}{2} \, , \\
    & && R_{h \pi_3 \pi_3 h} \vbar_{\mathcal{H}} &&= - \frac{K' (F_C' + F_T')}{4K} - \frac{(F_C' + F_T')^2}{4 (F_C + F_T)} + \frac{F_C'' + F_T''}{2} \, , \\
    & && R_{\pi_1 \pi_2 \pi_2 \pi_1} \vbar_{\mathcal{H}} &&= - \frac{F_C}{v^2} + \frac{F_C'^2}{4 K} + \frac{3 F_T}{v^2} \, , \\
    & R_{\pi_1 \pi_3 \pi_3 \pi_1} \vbar_{\mathcal{H}} & \; = \; & R_{\pi_2 \pi_3 \pi_3 \pi_2} \vbar_{\mathcal{H}} &&= - \frac{F_C}{v^2} + \frac{F_C'^2}{4 K} - \frac{2 F_T}{v^2} + \frac{F_C' \, F_T'}{4K} - \frac{F_T^2}{v^2 F_C} \, ,
\end{alignat}
and a non-independent curvature component
\begin{equation}
    R_{\pi_1 \pi_2 \pi_3 h} \vbar_{\mathcal{H}} = - \frac{F_C' \, F_T^{\vphantom{'}}}{v F_C} + \frac{F_T'}{v} \, .
\end{equation}

Next is the fermion vector bundle $\mathcal{E}$, which is diagonal in chirality in the sense that any mixed component of a geometric quantity between the two $Q_L$ and $Q_R$ doublets vanishes. We continue to treat $\pi_A$ as small. The leading terms of the connection coefficients in $h$ are
\begin{align}
    \tensor{\Gamma}{^{t_L}_{h t_L}} &\supset \frac{1}{2 k_L^+} \left ( {k_L^+}' + \frac{2}{v} \, \omega_L^+ \right ) \, , \hspace{3.85em} \tensor{\Gamma}{^{t_R}_{h t_R}} \supset \frac{1}{2 k_R^+} \left ( {k_R^+}' + \frac{2}{v} \, \omega_R^+ \right ) \, , \\
    \tensor{\Gamma}{^{b_L}_{h b_L}} &\supset \frac{1}{2 k_L^-} \left ( {k_L^-}' + \frac{2}{v} \, \omega_L^- \right ) \, , \hspace{3.75em} \tensor{\Gamma}{^{b_R}_{h b_R}} \supset \frac{1}{2 k_R^-} \left ( {k_R^-}' + \frac{2}{v} \, \omega_R^- \right ) \, , \\
    \tensor{\Gamma}{^{t_L}_{h b_L}} &\supset \left [ \frac{k_{LT}}{2 k_L^+ k_L^-} \left ( {k_L^-}' + \frac{2}{v} \, \omega_L^- \right ) - \frac{1}{2 k_L^+} \left ( k_{LT}' + \frac{2}{v} \, \omega_{LT} \right ) \right ] \left ( \zeta_1 - i \zeta_2 \right ) \, , \\
    \tensor{\Gamma}{^{b_L}_{h t_L}} &\supset \left [ \frac{k_{LT}}{2 k_L^+ k_L^-} \left ( {k_L^+}' + \frac{2}{v} \, \omega_L^+ \right ) - \frac{1}{2 k_L^-} \left ( k_{LT}' + \frac{2}{v} \, \omega_{LT} \right ) \right ] \left ( \zeta_1 + i \zeta_2 \right ) \, , \\[0.5em]
    \tensor{\Gamma}{^{t_R}_{h b_R}} &= 0 \, , \hspace{12em} \tensor{\Gamma}{^{b_R}_{h t_R}} = 0 \, ,
\end{align}
while those in $\pi_A$ are
\begingroup
\allowdisplaybreaks
\begin{align}
    \tensor{\Gamma}{^{t_L}_{\pi_1 t_L}} &\supset - \frac{k_{LT}}{2 v k_L^+ k_L^-} \left ( 2 \varpi_L - i k_{LT} - 2 i \varpi_{LT1} \right ) \left ( \zeta_1 - i \zeta_2 \right ) \\
    &\quad - \frac{1}{v k_L^+} \left [ \left ( \varpi_L - \varpi_{LT2} - \varpi_{LT3} \right ) \zeta_1 - \left ( k_{LT} + 2 \varpi_{LT1} \right ) \zeta_2 \right ] \, , \notag \\
    \tensor{\Gamma}{^{t_L}_{\pi_2 t_L}} &\supset - \frac{i k_{LT}}{2 v k_L^+ k_L^-} \left ( 2 \varpi_L - i k_{LT} - 2 i \varpi_{LT1} \right ) \left ( \zeta_1 - i \zeta_2 \right ) \\
    &\quad - \frac{1}{v k_L^+} \left [ \left ( \varpi_L - \varpi_{LT2} - \varpi_{LT3} \right ) \zeta_2 + \left ( k_{LT} + 2 \varpi_{LT1} \right ) \zeta_1 \right ] \, , \notag \\
    \tensor{\Gamma}{^{t_L}_{\pi_3 t_L}} &\supset - \frac{1}{v k_L^+} \left ( \varpi_L + \varpi_{LT2} + \varpi_{LT3} \right ) \, , \\
    \tensor{\Gamma}{^{t_R}_{\pi_A t_R}} &\supset \frac{1}{v k_R^+} \left ( \varpi_R + \varpi_{RT2} + \varpi_{RT3} \right ) \xi_A \, , \\
    \tensor{\Gamma}{^{b_L}_{\pi_1 b_L}} &\supset - \frac{k_{LT}}{2 v k_L^+ k_L^-} \left ( 2 \varpi_L + i k_{LT} + 2 i \varpi_{LT1} \right ) \left ( \zeta_1 + i \zeta_2 \right ) \\
    &\quad + \frac{1}{v k_L^-} \left [ \left ( \varpi_L + \varpi_{LT2} - \varpi_{LT3} \right ) \zeta_1 - \left ( k_{LT} + 2 \varpi_{LT1} \right ) \zeta_2 \right ] \, , \notag \\
    \tensor{\Gamma}{^{b_L}_{\pi_2 b_L}} &\supset \frac{i k_{LT}}{2 v k_L^+ k_L^-} \left ( 2 \varpi_L + i k_{LT} + 2 i \varpi_{LT1} \right ) \left ( \zeta_1 + i \zeta_2 \right ) \\
    &\quad + \frac{1}{v k_L^-} \left [ \left ( \varpi_L + \varpi_{LT2} - \varpi_{LT3} \right ) \zeta_2 + \left ( k_{LT} + 2 \varpi_{LT1} \right ) \zeta_1 \right ] \, , \notag \\
    \tensor{\Gamma}{^{b_L}_{\pi_3 b_L}} &\supset \frac{1}{v k_L^-} \left ( \varpi_L - \varpi_{LT2} + \varpi_{LT3} \right ) \, , \\
    \tensor{\Gamma}{^{b_R}_{\pi_A b_R}} &\supset - \frac{1}{v k_R^-} \left ( \varpi_R - \varpi_{RT2} + \varpi_{RT3} \right ) \xi_A \, , \\
    \tensor{\Gamma}{^{t_L}_{\pi_1 b_L}} &\supset - \frac{1}{2 v k_L^+} \left ( 2 \varpi_L + i k_{LT} + 2 i \varpi_{LT1} \right ) \, , \\
    \tensor{\Gamma}{^{b_L}_{\pi_1 t_L}} &\supset - \frac{1}{2 v k_L^-} \left ( 2 \varpi_L - i k_{LT} - 2 i \varpi_{LT1} \right ) \, , \\
    \tensor{\Gamma}{^{t_L}_{\pi_2 b_L}} &\supset \frac{i}{2 v k_L^+} \left ( 2 \varpi_L + i k_{LT} + 2 i \varpi_{LT1} \right ) \, , \\
    \tensor{\Gamma}{^{b_L}_{\pi_2 t_L}} &\supset - \frac{i}{2 v k_L^-} \left ( 2 \varpi_L - i k_{LT} - 2 i \varpi_{LT1} \right ) \, , \\
    \tensor{\Gamma}{^{t_L}_{\pi_3 b_L}} &\supset \bigg [ \frac{k_{LT}}{v k_L^+ k_L^-} \left ( \varpi_L - \varpi_{LT2} + \varpi_{LT3} \right ) \\
    &\qquad + \frac{1}{2 v k_L^+} \left ( 2 \varpi_L - i k_{LT} - 2 i \varpi_{LT1} + 4 \varpi_{LT3} \right ) \bigg ] \left ( \zeta_1 - i \zeta_2 \right ) \, , \notag \\
    \tensor{\Gamma}{^{b_L}_{\pi_3 t_L}} &\supset \bigg [ - \frac{k_{LT}}{v k_L^+ k_L^-} \left ( \varpi_L + \varpi_{LT2} + \varpi_{LT3} \right ) \\
    &\qquad + \frac{1}{2 v k_L^-} \left ( 2 \varpi_L + i k_{LT} + 2 i \varpi_{LT1} + 4 \varpi_{LT3} \right ) \bigg ] \left ( \zeta_1 + i \zeta_2 \right ) \, , \notag \\
    \tensor{\Gamma}{^{t_R}_{\pi_1 b_R}} &\supset - \frac{1}{v k_R^+} \left ( \varpi_R + i \varpi_{RT1} \right ) \, , \\
    \tensor{\Gamma}{^{b_R}_{\pi_1 t_R}} &\supset - \frac{1}{v k_R^-} \left ( \varpi_R - i \varpi_{RT1} \right ) \, , \\
    \tensor{\Gamma}{^{t_R}_{\pi_2 b_R}} &\supset \frac{i}{v k_R^+} \left ( \varpi_R + i \varpi_{RT1} \right ) \, , \\
    \tensor{\Gamma}{^{b_R}_{\pi_2 t_R}} &\supset - \frac{i}{v k_R^-} \left ( \varpi_R - i \varpi_{RT1} \right ) \, , \\
    \tensor{\Gamma}{^{t_R}_{\pi_3 b_R}} &\supset - \frac{1}{v k_R^+} \left ( \varpi_R + i \varpi_{RT1} \right ) \left ( \zeta_1 - i \zeta_2 \right ) \, , \\
    \tensor{\Gamma}{^{b_R}_{\pi_3 t_R}} &\supset - \frac{1}{v k_R^-} \left ( \varpi_R - i \varpi_{RT1} \right ) \left ( \zeta_1 + i \zeta_2 \right ) \, .
\end{align}%
\endgroup
The coefficients for the conjugate connection are complex conjugates of the above. We obtain the two sets of two-fermion curvature components
\begin{align}
    R_{\bar{t}_L t_L h \pi_3} \vbar &= \frac{{k_L^+}'}{v k_L^+} \left ( \varpi_L + \varpi_{LT2} + \varpi_{LT3} \right ) - \frac{1}{v} \left ( \varpi_L' + \varpi_{LT2}' + \varpi_{LT3}' \right ) \, , \\
    R_{\bar{t}_R t_R h \pi_3} \vbar &= \frac{{k_R^+}'}{v k_R^+} \left ( \varpi_R + \varpi_{RT2} + \varpi_{RT3} \right ) - \frac{1}{v} \left ( \varpi_R' + \varpi_{RT2}' + \varpi_{RT3}' \right ) \, , \\
    R_{\bar{b}_L b_L h \pi_3} \vbar &= - \frac{{k_L^-}'}{v k_L^-} \left ( \varpi_L - \varpi_{LT2} + \varpi_{LT3} \right ) + \frac{1}{v} \left ( \varpi_L' - \varpi_{LT2}' + \varpi_{LT3}' \right ) \, , \\
    R_{\bar{b}_R b_R h \pi_3} \vbar &= - \frac{{k_R^-}'}{v k_R^-} \left ( \varpi_R - \varpi_{RT2} + \varpi_{RT3} \right ) + \frac{1}{v} \left ( \varpi_R' - \varpi_{RT2}' + \varpi_{RT3}' \right ) \, , \\
    R_{\bar{t}_L t_L \pi_1 \pi_2} \vbar &= \frac{2}{v^2} \left ( \varpi_L - \varpi_{LT2} - \varpi_{LT3} \right ) + \frac{i}{2 v^2 k_L^-} \left [ \left ( 2 \varpi_L - i k_{LT} \right )^2 + 4 \varpi_{LT1}^2 \right ] \, , \\
    R_{\bar{t}_R t_R \pi_1 \pi_2} \vbar &= - \frac{2}{v^2} \left ( \varpi_R + \varpi_{RT2} + \varpi_{RT3} \right ) + \frac{i}{2v^2 k_R^-} \left [ 4 \varpi_R^2 + 4 \varpi_{RT1}^2 \right ] \, , \\
    R_{\bar{b}_L b_L \pi_1 \pi_2} \vbar &= - \frac{2}{v^2} \left ( \varpi_L + \varpi_{LT2} - \varpi_{LT3} \right ) - \frac{i}{2 v^2 k_L^+} \left [ \left ( 2 \varpi_L + i k_{LT} \right )^2 + 4 \varpi_{LT1}^2 \right ] \, , \\
    R_{\bar{b}_R b_R \pi_1 \pi_2} \vbar &= \frac{2}{v^2} \left ( \varpi_R - \varpi_{RT2} + \varpi_{RT3} \right ) - \frac{i}{2 v^2 k_R^+} \left [ 4 \varpi_R^2 + 4 \varpi_{RT1}^2 \right ] \, ,
\end{align}
and
\begingroup
\allowdisplaybreaks
\begin{align}
    R_{\bar{t}_L b_L h \pi_1} \vbar &= i R_{\bar{t}_L b_L h \pi_2} \vbar = - R_{\bar{b}_L t_L h \pi_1}^* \vbar = - i R_{\bar{b}_L t_L h \pi_2}^* \vbar \\
    &= \frac{i}{v^2} \, \omega_{LT} - \frac{1}{v} \left ( \varpi_L' + i \varpi_{LT1}' \right ) + \frac{1}{4 v k_L^+} \left ( {k_L^+}' - \frac{2}{v} \, \omega_L^+ \right ) \left ( 2 \varpi_L + i k_{LT} + 2 i \varpi_{LT1} \right ) \notag \\
    &\quad + \frac{1}{4 v k_L^-} \left ( {k_L^-}' + \frac{2}{v} \, \omega_L^- \right ) \left ( 2 \varpi_L - i k_{LT} + 2 i \varpi_{LT1} \right ) \notag \, , \\
    R_{\bar{t}_R b_R h \pi_1} \vbar &= i R_{\bar{t}_R b_R h \pi_2} \vbar = - R_{\bar{b}_R t_R h \pi_1}^* \vbar = - i R_{\bar{b}_R t_R h \pi_2}^* \vbar \\
    &= - \frac{1}{v} \left ( \varpi_R' + i \varpi_{RT1}' \right ) \notag \\
    &\quad + \frac{1}{4v} \left [ \frac{1}{k_R^+} \left ( {k_R^+}' - \frac{2}{v} \, \omega_R^+ \right ) + \frac{1}{k_R^-} \left ( {k_R^-}' + \frac{2}{v} \, \omega_R^- \right ) \right ] \left ( 2 \varpi_R + 2 i \varpi_{RT1} \right ) \notag \, , \\
    R_{\bar{t}_L b_L \pi_1 \pi_3} \vbar &= i R_{\bar{t}_L b_L \pi_2 \pi_3} \vbar = - R_{\bar{b}_L t_L \pi_1 \pi_3}^* \vbar = - i R_{\bar{b}_L t_L \pi_2 \pi_3}^* \vbar \\
    &= \frac{2i}{v^2} \left ( \varpi_L + \varpi_{LT3} \right ) - \frac{1}{2 v^2 k_L^+} \left ( \varpi_L + \varpi_{LT2} + \varpi_{LT3} \right ) \left ( 2 \varpi_L + i k_{LT} + 2 i \varpi_{LT1} \right ) \notag \\
    &\quad - \frac{1}{2 v^2 k_L^-} \left ( \varpi_L - \varpi_{LT2} + \varpi_{LT3} \right ) \left ( 2 \varpi_L - i k_{LT} + 2 i \varpi_{LT1} \right ) \, , \notag \\
    R_{\bar{t}_R b_R \pi_1 \pi_3} \vbar &= i R_{\bar{t}_R b_R \pi_2 \pi_3} \vbar = - R_{\bar{b}_R t_R \pi_1 \pi_3}^* \vbar = - i R_{\bar{b}_R t_R \pi_2 \pi_3}^* \vbar \\
    &= - \frac{2i}{v^2} \left ( \varpi_R + i \varpi_{RT1} \right ) - \frac{1}{2v^2} \bigg [ \frac{1}{k_R^+} \left ( \varpi_R + \varpi_{RT2} + \varpi_{RT3} \right ) \notag \\
    &\hspace{13em} + \frac{1}{k_R^-} \left ( \varpi_R - \varpi_{RT2} + \varpi_{RT3} \right ) \bigg ] \left ( 2 \varpi_R + 2 i \varpi_{RT1} \right ) \, , \notag 
\end{align}%
\endgroup
that are non-vanishing at the vacuum.

We do not distinguish the $\Gamma$'s and $R$'s of the graded Riemannian metric $G$ from their underlying counterparts on $\mathcal{M}$ and $\mathcal{E}$, which differ only by $\mathcal{O}(\chi^2)$ corrections.

\bibliographystyle{JHEP}
\bibliography{biblio.bib}

\end{document}